# Selective Enrichment of Full AAV Capsids

Vivekananda Bal[1], Jacqueline M. Wolfrum[2], Paul W. Barone[2], Stacy L. Springs[2], Anthony J. Sinskey[2,3], Robert M. Kotin[2,4], and Richard D. Braatz[1,2]

[1] Department of Chemical Engineering, Massachusetts Institute of Technology, Cambridge, MA, USA

[2] Center for Biomedical Innovation, Massachusetts Institute of Technology, Cambridge, MA, USA

[3] Department of Biology, Massachusetts Institute of Technology, Cambridge, MA, USA

[4] Gene Therapy Center, University of Massachusetts Chan Medical School, Worcester, MA, USA

**Gene therapies using recombinant adeno-associated virus (rAAV) have been developed to treat monogenic and acquired diseases but are currently the most expensive drugs due, in part, to high manufacturing costs. The cells producing rAAV generate substantial quantities of empty (50-90%) and partially filled (indeterminant) capsids that must be removed prior to final formulation. The conventional separation processes are inefficient in removing empty and partially filled capsids, have low yield, scale poorly, expensive, time consuming and need additional purification steps. This article demonstrates one step separation of full capsids from a mixture of full, partially filled, and empty capsids, and other protein impurities using selective crystallization – a purification process, which is first time in protein purification and is performed without physically/chemically modifying the target component (full capsid) for the first time in the history of selective crystallization, and is highly-efficient, highly-scalable, economical, and improves product quality. Hanging-drop vapor diffusion experiments were used to scout crystallization conditions in which full and empty capsids crystallize, then to define conditions in which crystals of full, empty, or both full and empty capsids nucleate and grow. The experimental results for rAAV serotypes 5, 8, and 9 as exemplary vectors and scale-up results show that full capsids can be selectively crystallized and separated in one step from a mixture of full, partially filled, and empty capsids, and other protein and salt impurities with full capsid enrichment of > 80%, approximately 20% higher, and yield of > 90%, approximately 30% higher from the existing methods, keeping their biological activity intact, in a short period of time (< 4 h), with approximately 87% reduction in processing time from the existing processing time and without the need of additional purification steps and in one round.**

Recombinant adeno-associated virus (rAAVs) gene therapy has demonstrated clinical improvement for patients affected by congenital blindness (Leber's congenital amaurosis), hemophilia A, hemophilia B, aromatic amino acid deficiency, and spinal muscular atrophy.[1] > 70% of the gene therapies currently approved by drug regulatory agencies globally are viral vector based.[2–4] The largest class of these gene therapies are based on rAAVs, which are the focus of > 200 clinical trials.[3]

Depending on the transgene, capsid, and manufacturing platform, the rAAV particles may initially contain 50% to 90% empty capsids (particles that do not contain the therapeutic transgene) (**Fig. 1a**) and an indeterminant percentage of partially filled capsids.[5–7] The empty capsids, provide no therapeutic benefit, increase the capsid antigen load, reduce the transduction efficiency, and contribute to immune responses such as the thrombotic microangiopathies.[8–11] These observations have stimulated the development of methods for removing empty capsids to improve the efficacy and safety of the therapy. However, separating empty and particularly partially filled capsids from full capsids is extremely challenging, due to their very similar physical properties, such as size, density (1.41 gm/cm$^3$ for full and 1.31 gm/cm$^3$ for empty, assuming 4.7 kbp virion genome), and isoelectric point (5.9 for full and 6.3 for empty).[12,13] Chromatography and ultracentrifugation are widely used to purify/enrich full particles (**Fig. 1b**).[10,14,15] Ultracentrifugation (UC) exploits the difference in buoyant density and ion-exchange chromatography (IEX) exploits the difference in particle charge density between full and empty capsids. But these conventional methods do not effectively remove empty and partially filled capsids. Additionally, they are poorly scalable and incur high costs, and involve long processing times, additional manipulations, and higher product losses which make them unsuitable for

use in commercial clinical-grade vector production.[16–18]

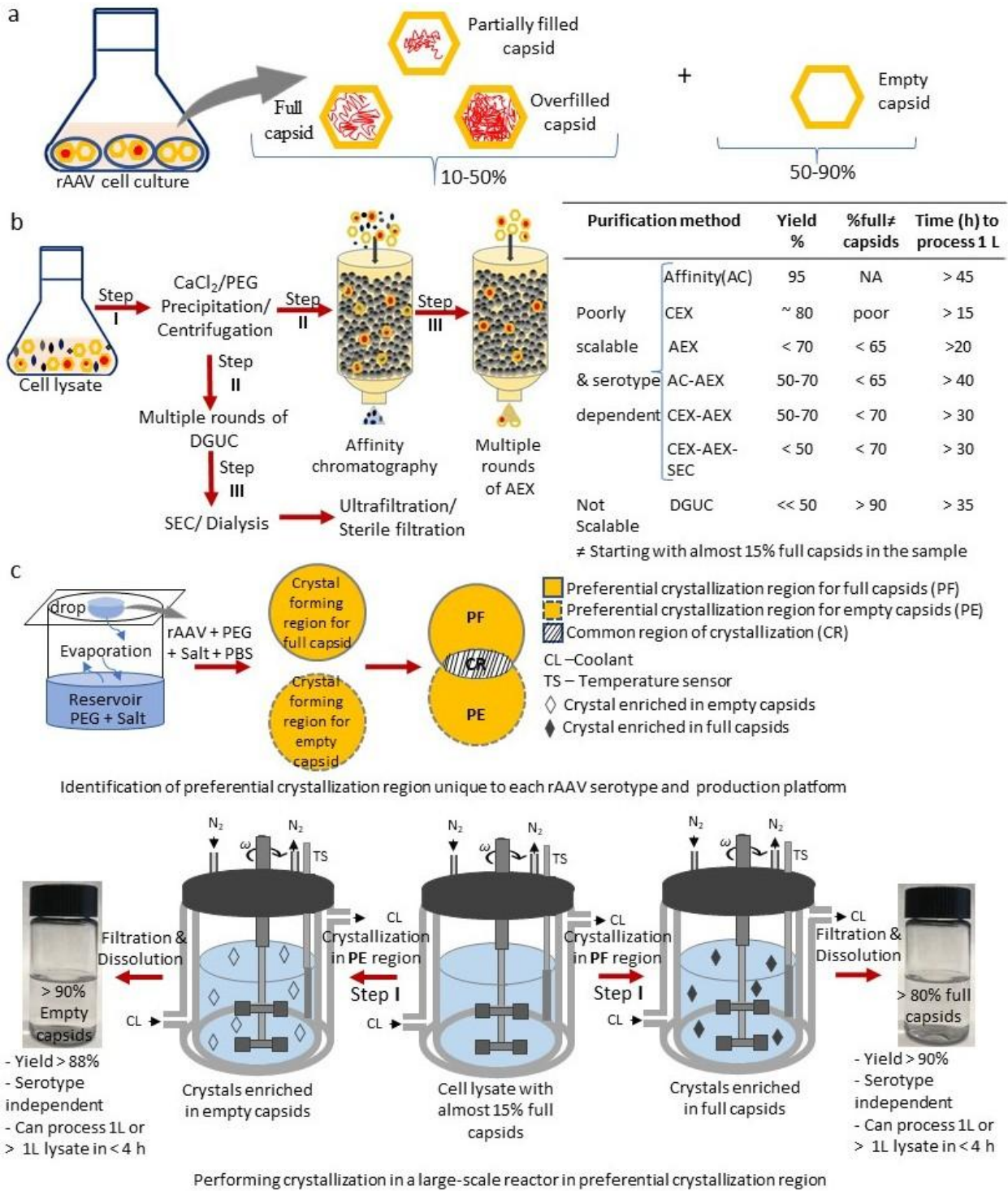

**Fig.1:** Schematic of purification/separation of either full or empty capsids from their mixture/cell lysate. (a) Production of rAAVs in cells suspension, (b) Schematic of industrial scale purification of full capsids with table showing the percentage yield/recovery, enrichment achieved (% full capsids), and time required to process 1 L of cell lysate material, (c) Proposed idea of selective crystallization: Identification of preferential crystallization region unique to each serotype and production platform for full and empty capsids and performing preferential crystallization experiment to separate full and empty capsids from the cell lysate. Captions showing the enrichment (% full capsids) and yield achieved after preferential crystallization and the corresponding processing time for 1 L or > 1 L cell lysate

Here we propose that the slight differences in charge density (Supplementary **Figs. 1 and 2**) are sufficient for preferential crystallization of full rAAV capsids. The process provides a scalable and economical method for rAAV capsid purification compatible with commercial industrial scale used in conventional purification methods. Though, preferential crystallization is common in small molecule pharmaceutical industry, it has not been used in protein purification previously and this is the first-time preferential crystallization has been performed without physically/chemically modifying the target component (i.e., full capsids in this case). The salt, pH, and organic precipitant provide "levers" that may be manipulated to influence the charge density surrounding each capsid, and the charged interactions between the full, partially filled, and empty capsids. The solution pH has a strong effect on capsid surface charge density, whereas salt influences the charge environment around capsid by controlling the electrical double-layer thickness/zeta-potential, just as for other proteins (Supplementary **Figs. 1 and 2**).[19–23] As such, salt concentration controls the rate of crystallization by shielding capsid's inherent charge density, just as for protein molecules.[24–30] Together, pH and salt concentration thus directly control the capsid-to-capsid interactions and hence the solubility and crystallization rates of full, partially filled, and empty capsids.

In this article, we identify solution conditions for the preferential crystallization of full and empty capsids of AAV serotypes 5, 8, and 9 as exemplary vectors. To establish a higher resolution "picture" of crystallization conditions, a wide range of precipitant and pH concentrations were screened in a hanging-drop vapor diffusion system (Supplementary **Figs. 3, 4**) and a phase diagram is constructed for both "full" and "empty" capsid samples (**Fig. 1c and** Supplementary **Fig. 4**). To identify the preferential crystallization conditions, phase diagrams for both "full" and "empty" capsids were plotted in the same figure. This clearly shows the overlapped region, where both full and empty capsids co-crystallize, and the preferential crystallization regions for full capsids and for empty capsids (**Fig. 1c**). Once the preferential regions are identified, crystallization experiments were performed in that region to separate full capsids, which were redissolved for further use. These preferential crystallization regions are unique to each serotype and production platform (e.g., sf9 or HEK293) and can thus be used to preferentially crystallize full capsids from the cell lysate without the need of generating the preferential crystallization conditions afresh as long as the serotype and the production platform remain the same.

PEG and NaCl are the most commonly used precipitants for protein crystallization.[31–33] The NaCl influences the proteins solubility by "salting in" and "salting out" effects, whereas the PEG functions as an aquacide effectively concentrating the proteins.[34]

**Figs. 2a-d (column 1 and 2)** shows the corresponding phase diagrams for "full" and "empty" capsids of rAAV5, rAAV9, and rAAV8, respectively, as a function of pH, PEG8000 and NaCl concentrations. The graphical representation reveals that crystallization occurs within a closed and bounded region. Empty and full capsids of different serotypes show appreciably different solution behavior depending on pH and NaCl and PEG8000 concentrations. Higher charge density on the full capsids compared to the empty capsids for rAAV serotypes 5 and 9 tend to increase the solubility and this causes the crystallization of full capsids to occur at relatively higher PEG and NaCl concentrations. In contrary, significant post-translational modifications (PTMs) in empty capsids compared to full capsids for rAAV8 (Supplementary **Fig. 14b**) tend to increase solubility for empty capsids (as observed in other molecules or compounds[35–37]) and this allows the empty capsids to crystallize at relatively higher PEG and NaCl concentrations.

Regions, where either full or empty rAAV particles can be preferentially crystallized and the common regions, where both empty and filled capsids co-crystallize, are clearly observed in **Figs. 2a-c column 3** for **best-case conditions** and in Supplementary **Fig. 5** for each pH value tested. The pH values in these experiments were chosen to be near the isoelectric point of full and empty rAAV, i.e., where the net charge on the surface is zero (i.e., zero zeta-potential, Supplementary **Fig. 2**) and the capsids have the lowest solubility just as for other proteins.[38–41] These preferential regions depend on pH and PEG and NaCl concentrations and the net charge of the assembled coat proteins and the DNA within the capsid.

Additionally, other precipitants were evaluated for differential crystallization of AAV capsids. Crystallization conditions were screened using PEG6000 and

NaCl, as well as PEG8000 and MgCl$_2$ (**Fig. 2d**) as precipitants. Both cases show that "full" and "empty" capsids can be preferentially crystallized from a mixture. As such, the preferential crystallization was not sensitive to the molecular weight of PEG or the valency of inorganic salt.

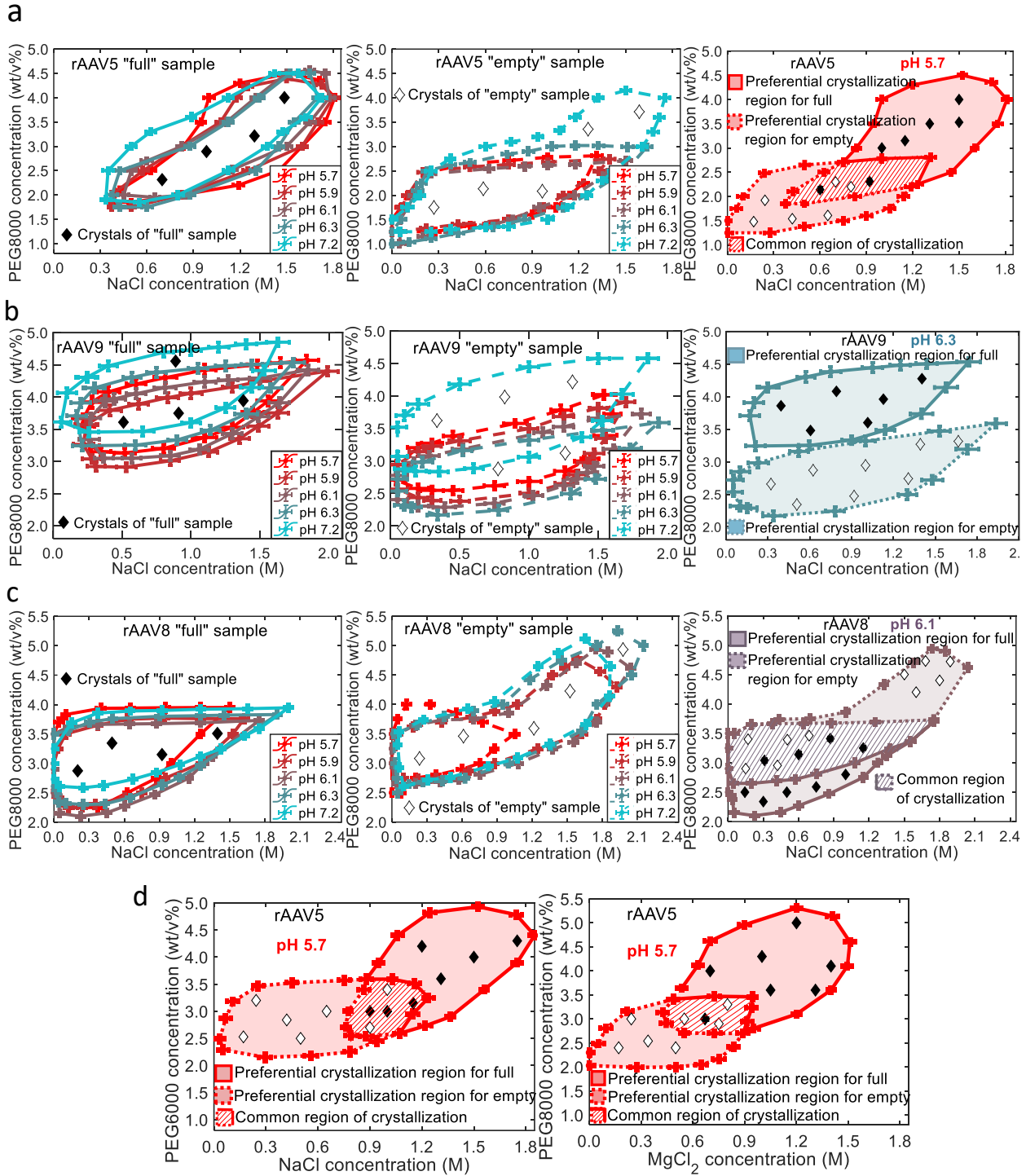

**Fig.2:** Identification of preferential crystallization region for full and empty capsids of Sf9 produced rAAV. (a) Phase diagram for rAAV5 at different pH for NaCl and PEG8000 as a precipitant. Column 1: phase diagram for full capsids, Column 2: phase diagram for empty capsids, Column 3: preferential crystallization region for full capsids and for empty capsids of rAAV5 at Ph 5.7 as an example. (b) Phase diagram for rAAV9 as a function of pH for NaCl and

PEG8000 as a precipitant. Column 1: phase diagram for full capsids, Column 2: phase diagram for empty capsids, Column 3: preferential crystallization region for full and empty capsids of rAAV9 at Ph 6.3 as an example. (c) Phase diagram for rAAV8 at different pH for NaCl and PEG8000 as a precipitant. Column 1: phase diagram for full capsids, Column 2: phase diagram for empty capsids, Column 3: preferential crystallization region for full and empty capsids of rAAV8 at pH 6.1 as an example. (d) Effect of precipitant on the preferential crystallization regime. left: effect of PEG6000 on the preferential crystallization of full and empty rAAV5 capsids at pH 5.7, right: effect of $MgCl_2$ on the preferential crystallization of full and empty capsids of rAAV5 at pH 5.7. In all cases, length of the transgene was 2.46 kbp.

A solubility analysis shows that the visible particles are composed of capsids rather than NaCl or PEG (Supplementary **Fig. 6**). To determine whether co-formulation of NaCl and PEG affects the solubility of either component,[42–45] a set of control experiments was performed for the combinations across the full ranges of PEG concentration (0.5 to 8 wt/v%) and NaCl concentration (0.005 to 2.5 M) in the absence of rAAV capsids. No crystals were observed under these conditions, confirming that rAAV capsids were essential for crystallization. A polarization study (Supplementary **Figs. 7 and 8**) shows that the particles are crystalline, which is further supported by SEM imaging (Supplementary **Fig. 9**).

To evaluate quantitatively the enrichment of full or empty capsids after preferential crystallization, ddPCR and ELISA experiments were performed on dissolved crystals, and the percentage of full and empty capsids was calculated for each experimental condition (supplementary **Fig. 10**). The corresponding average value for each starting capsid concentration is shown in **Fig. 3a Tables 2** and **3**. Conditions resulting in preferential crystallization of full capsids results in higher fraction of full capsids in crystals regardless of the fraction of full capsids in the starting sample. Significantly, 74-80% full capsid enrichment can be achieved starting with ~ 20% full capsids (**Fig. 3a Table 2**) by preferential crystallization. Obtaining greater enrichment is possible if the starting fraction of full capsids is higher. E.g., a starting sample with about 80% full capsids results in >94% full capsids in the crystals. Likewise, preferential crystallization of empty capsids (**Table 3** in **Fig. 3a**) results in crystals enriched with 71–78% empty capsids starting with ~20% empty capsids. Similar analysis was performed for rAAV9 and rAAV8 crystals. Here, too preferential crystallization produces crystals enriched in ~70–75% full capsids starting with about 19% full capsids. Thus, crystallization in the preferential region of full capsids selectively removed full capsids from the solution and in the preferential region of empty capsids selectively removed empty capsids from the solution.

To investigate the morphology of capsid after crystallization, TEM images of the crystallization-purified sample (Supplementary **Figs. 11fhjl**) were obtained. The micrographic images indicate that capsids are homogeneous in shape and size with a diameter of ~22–25 nm. Comparison of TEM image of crystallization-purified sample with that of reference sample (Supplementary **Figs. 11a-d**) shows that capsids morphology remains unaffected by the crystallization process. Although not designed as a quantitative assay, the percentage of full and empty capsids after preferential crystallization were calculated based on a total of ~2500 capsid particles from multiple TEM fields. This analysis suggests that, the full capsid preferential crystallization-purified sample (Supplementary **Figs. 11fj**) contains > 85% full capsids as compared to 20–25% full capsids in the starting sample (Supple-mentary **Figs. 11ei**). Likewise, empty capsid prefer-ential crystallization results in >90% empty capsids (Supplementary **Figs. 11hl**) from a starting sample of 20–25% empty capsids (Supplementary **Figs. 11gk**). Thus, TEM image analysis also supports the full-empty percentage calculated based on ddPCR and ELISA experiments

The ELISA, ddPCR, and TEM results are further supported by mass photometry results (Supplementary **Fig. 12**), which shows that a starting sample of rAAV5 of ~64% full + overfilled capsids is enriched to about 93% full; a starting sample of rAAV9 of ~74% full is enriched to about 97% full.

a

Table 1: % full capsids in crystals obtained after full capsid's preferential crystallization

| Serotype | % full capsids in stating sample | | |
|---|---|---|---|
| | 79±3.5 | 50±2.5 | 15.1±2.8 |
| rAAV5 | 97±3 | 90±4.5 | 81±2.5 |
| rAAV9 | 98±2.5 | 89±4 | 82±3.4 |
| rAAV8 | - | - | 82±2.1 |

Table 2: % empty capsids in crystals obtained after empty capsid's preferential crystallization

| Serotype | % empty capsids in stating sample | | |
|---|---|---|---|
| | 79±3.5 | 50±2.5 | 15±2.4 |
| rAAV5 | 97.5±5 | 90±4 | 78±2.5 |
| rAAV9 | 96.5±4 | 89±3 | 79.5±4 |
| rAAV8 | - | - | 81.9±1.5 |

b

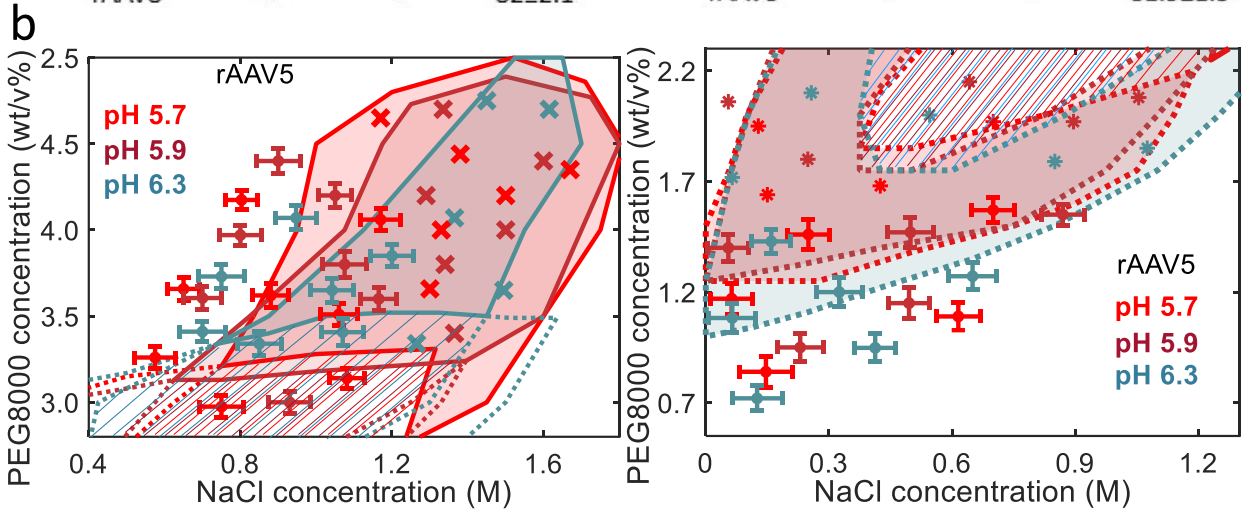

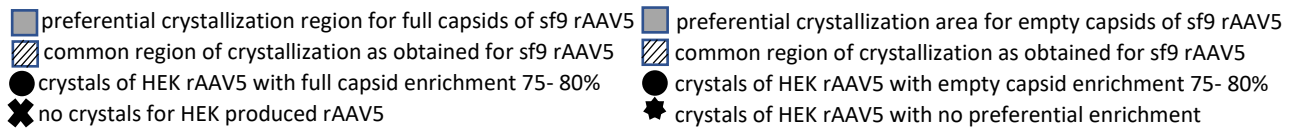

c

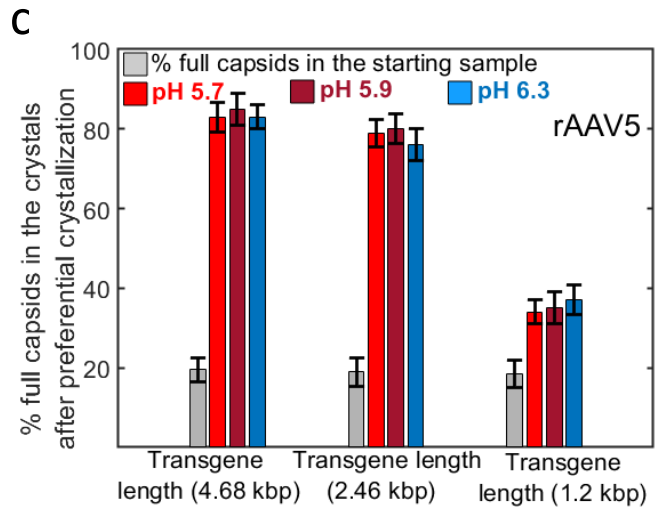

**Fig.3:** Percentage of full and empty capsids in the crystals obtained after preferential crystallization as measured by ddPCR and capsid ELISA experiment. (a) **Table 1**: show the average values and standard deviation of percentages of full capsids in the crystals obtained after crystallization in different conditions (as shown in Supplementary **Fig. 10**) in the preferential crystallization region of full capsids of Sf9 produced rAAV5, rAAV9, and rAAV8 for starting

samples with different fraction of full capsids. **Table 2**: show the average values and standard deviation of percentages of empty capsids in the crystals obtained after crystallization in different conditions (as shown in Supplementary **Fig. 10**) in the preferential crystallization region of empty capsids of Sf9 produced rAAV5, rAAV9, and rAAV8 for starting samples with different fraction of empty capsids. Full capsids used in these experiments contain transgene of length 2.46 kbp. (b) Left figure: Comparison of preferential crystallization region of full capsids of HEK293 cell produced rAAV5 (as shown by the successful experimental data points with error bars) with that of the full capsids of Sf9 produced rAAV5 (as shown by the shaded area, Right figure: Comparison of preferential crystallization region of empty capsids of HEK293 cell produced rAAV5 (as shown by the successful experimental data points with error bars) with that of the empty capsids of Sf9 produced rAAV5 (as shown by the shaded area. (c) Bar plots showing the average and standard deviation of percentages of full capsids in the crystals obtained after crystallization in different conditions (as shown in Supplementary **Fig. 14**) in the preferential crystallization region of full capsids of Sf9 produced rAAV5 with transgene length 4.68 kbp and 1.2 kbp.

As reported in literature,[46,47] rAAV capsids may undergo PTMs. The extent of PTMs may be influenced by processing, e.g., time of harvest, or production platform, e.g., HEK293 or Sf9 cells.[48] Being extremely sensitive to surface exposed substances,[35] crystallization conditions may be influenced by PTMs. An experiment was performed with HEK293 cells produced rAAV5 capsids as an example to evaluate whether the conditions for preferential crystallization depends on the production methods. The corresponding experimental results shown in **Fig. 3b** suggests that here also, full and empty capsids can be preferentially crystallized from their mixture and the conditions for crystallization are slightly different from that obtained for Sf9 produced rAAV5 (**Fig. 2a**). LC-MS (Liquid chromatography-Mass spectrometry) was performed to compare the PTMs level in Sf9 and HEK293 cell produced rAAV capsids (Supplementary **Fig. 13**).

The data indicate that rAAV5 PTMs are similar but slight differences appear between Sf9 and HEK293 cell produced vectors. However, without evaluating multiple vector batches from the same and different sources, no conclusion can be drawn about capsid PTMs produced in Sf9 or HEK293 cells. Results suggest that even a small difference in PTMs level is probably sufficient to cause a notable change in preferential crystallization conditions as crystallization process is extremely sensitive to the capsid structure, conformations and solubility, which are influenced by the external PTMS level.[35–37,49,50] The enrichment of full or empty capsids can be 75–80% for a starting sample with around 19% full or empty capsids. Thus, the preferential crystallization method itself and the capsid enrichment are not influenced by the source/quality of rAAV or PTMS level.

The nucleic acid contributes to the vectors mass and net charge, therefore, to understand the effect of transgene length on preferential crystallization and the corresponding enrichment, an experiment was performed with rAAV5 capsids carrying transgene of length half (1.2 kbp) and twice (4.68 kbp) the size of the full-length, wild-type virion genome. Results (**Fig. 3c**) suggests that the enrichment of capsids with transgene 1.2 kbp is almost half of the enrichments of capsids with transgene 2.46 kbp and the enrichment/selectivity becomes independent of transgene length above a certain transgene size. This suggests that the preferential crystallization kinetics/selectivity is strongly dependent on transgene length. Thus, by maintaining the conditions in preferential crystallization region of full capsids, full capsids can be selectively separated from the empty and partially filled capsids, whose therapeutic efficiency is questionable.[51–53] The literature on commercial purification methods such as IEX and UC do not explicitly talk about the removal of partially filled capsids, probably because of the process inefficiency and difficulty associated with measurement.[14,15,54] High selectivity of full capsids is most probably due to the strong dipolar nature of overall charge because of the non-centric alignment of the DNA inside a capsid as observed in other DNA viruses.[55] This allows the faster alignment/arrangement of full capsids in orderly manner to form crystals resulting in higher crystallization propensity and higher selectivity. For partially filled capsids, weak dipolar nature reduces the propensity of crystallization and hence poor selectivity.

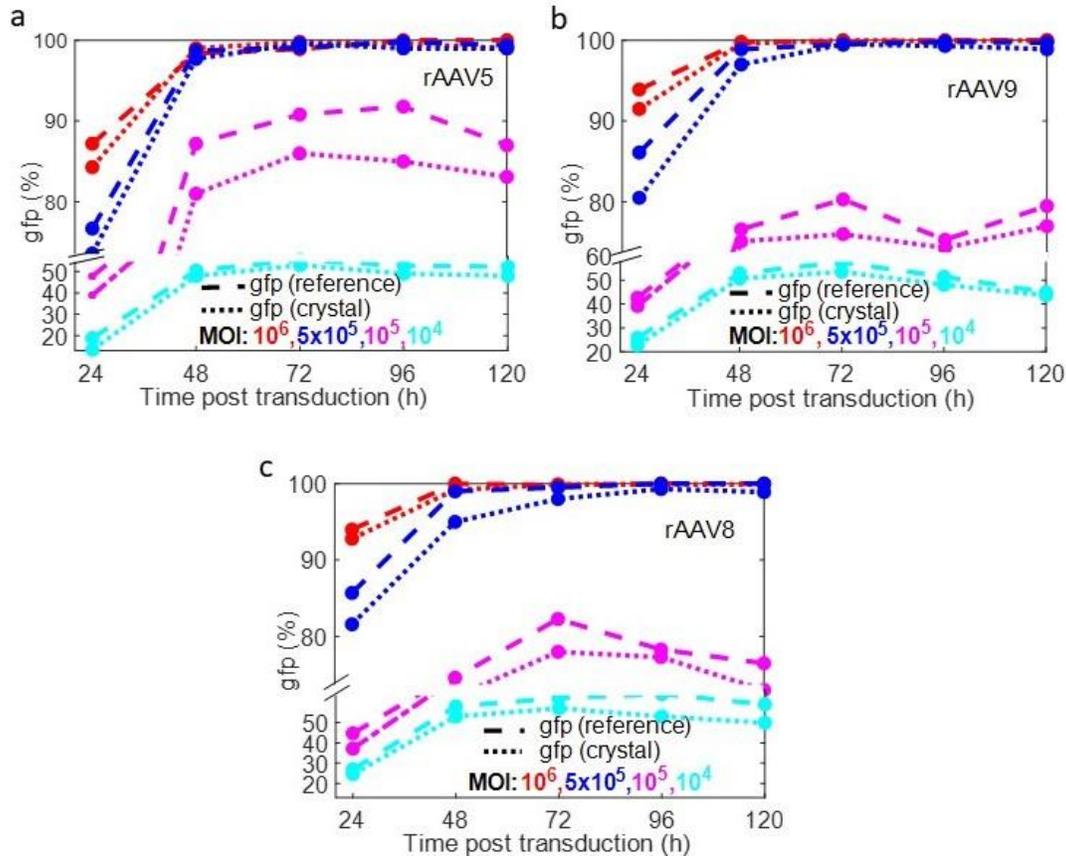

**Fig. 4:** Flow cytometry experiment results showing the variation of percentage of GFP-positive cells as a function of post transduction time for the transduction of HEK293 cell suspension with crystal stock solution as prepared from the crystals obtained after preferential crystallization of full capsids of Sf9 produced rAAV5 (a), rAAV9 (b), and rAAV8 (c) and transduction with the reference sample as purchased from Virovek. In all cases, cell viability remains well above 90% throughout five days (Supplementary **Fig. 15**). For all the serotypes, full capsid's transgene length was 2.46 kbp. Experimental conditions for preferential crystallization of full capsids: 4% PEG8000, 1.2 M NaCl at pH 5.7 for rAAV5 for a starting sample with 33% full capsids; 4% PEG8000, 0.6 M NaCl at pH 6.3 for rAAV9 for a starting sample with 37% full capsids; and 2.5% PEG8000, 0.6 M NaCl at pH 6.1 for rAAV8 for a starting sample with 41% full capsids.

To understand the effect of crystallization on biological activity, full rAAV5, rAAV8, and rAAV9 crystals were dissolved and used to transduce HEK293 cell suspension and analyzed for reporter gene, GFP, expression (**Fig. 4**) and viability (Supplementary **Fig. 15**) based on singlet cell population using flow cytometry every 24 h for 5 days. For all serotypes, viability and transduction efficiency (% GFP-positive cells) are comparable to that of the positive control (from Virovek) for all MOIs throughout the five days. In all cases, cell viability remains above 90% throughout the experiment ensuring that all the statistics are based on high population of live cells. It is observed that, for an MOI of $10^6$, > 85% of viable cells are transduced within 24 h, while only ~15% of viable cells are transduced in the same period of time when the virus dosage is reduced by an order of 2 (i.e., MOI $10^4$) and the transduction rate is so slow that only 40–50% of viable cells are transduced after day 5. These results demonstrate that biological activity of capsids remains preserved after crystallization and dissolution. Further HEK293T cells were transduced with dissolved crystal and cells were observed under fluorescence microscope every 24 h for five days and transduction efficiency was measured on day 5 (Supplementary **Fig. 16**). In nearly all cases, weak GFP fluorescence intensity appeared after 24 h, reaching maximum fluorescence intensity at day 4 with MOIs of $10^5$ and $10^6$ approaching the transduction limit. For all serotypes, transduction efficiency agrees well with that of the positive control (from Virovek), indicating that biological activity and stability of the full capsids were preserved during crystallization and after dissolution.

Transduction experiments, analogous to tissue culture infectious dose (TCID50) assays are performed to obtain a more precise value of biologically active vector titer. Biologically active vector titer for crystal stock solution was found to be close to that of positive control (Supplementary **Fig. 17 &** Supplementary **Table 1**), demon-strating that the biological activity of full capsids remains preserved even after crystallization.

SDS-PAGE gel electrophoresis was performed to assess vector purity and determine integrity of capsid proteins (**Figs. 5ab**) following crystallization. These images confirm the presence of all three VP proteins VP1, VP2, and VP3 in the crystallized capsids and the quantification of proteins indicates that the ratio of VPs in the crystallized sample remains nearly the same as before crystallization for all serotypes. The images also suggest that all the vector samples (from Virovek) contain low-molecular-weight protein impurities (**Fig. 5** and Supplementary **Fig. 18**), which are removed by the preferential crystallization. To confirm if the protein impurities/band at molecular weight 20–30 kbp and 40 kbp are fragments of VP proteins, western blot (**Figs. 5ab**) was performed and the results confirm that the impurities are not fragments of VP proteins. Thus, other protein impurities, which seem to be originated during cell lysis, and conventionally removed by affinity chromatography, are also removed by preferential crystallization eliminating the need of the former.

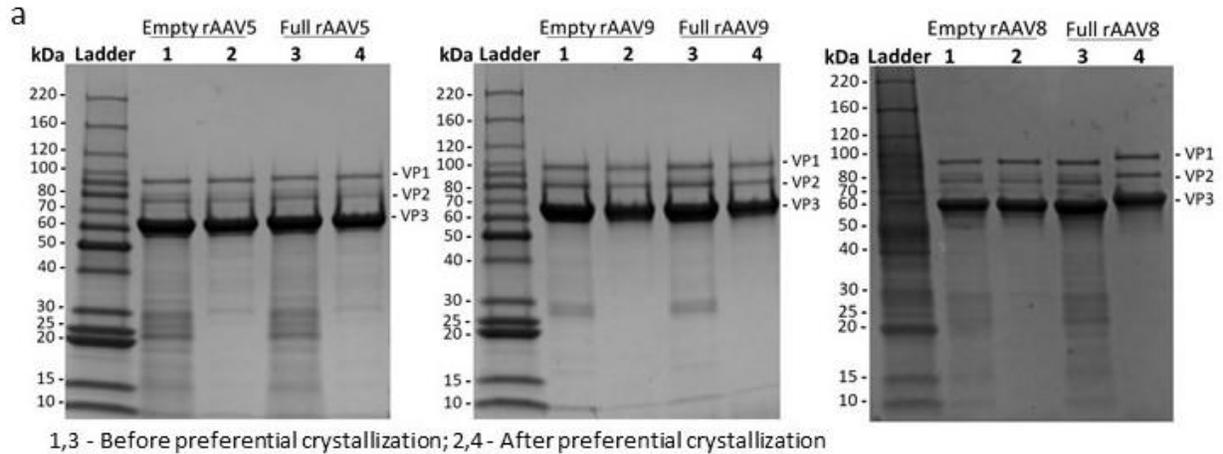

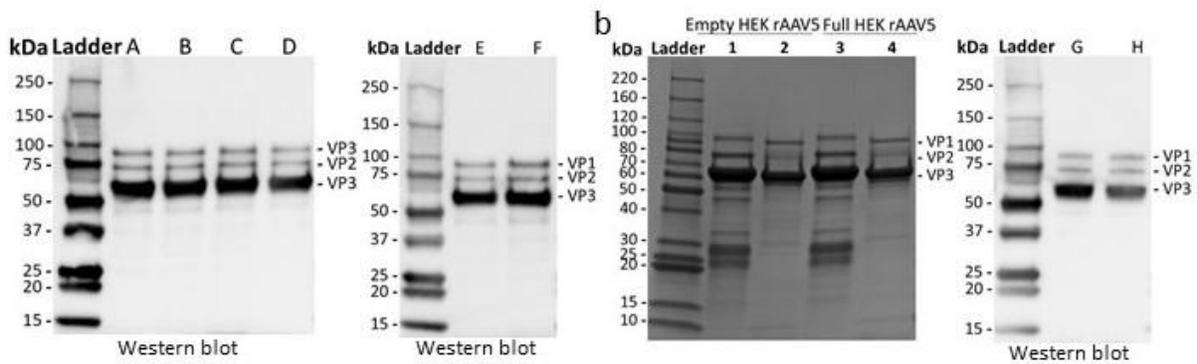

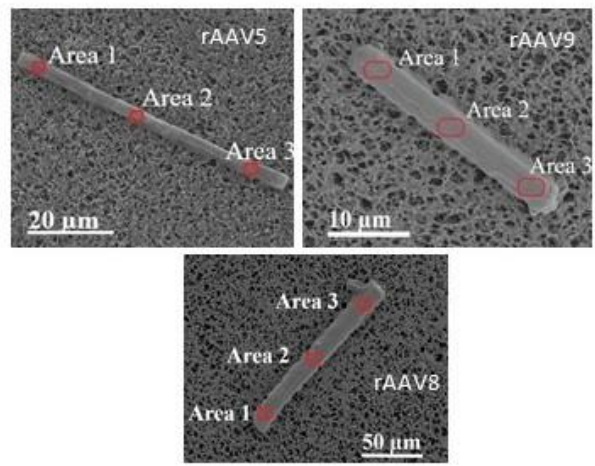

Table 3: Elemental composition of crystals as obtained in EDAX analysis

| Elements | rAAV5 (atomic%) | rAAV9 (atomic %) | rAAV8 (atomic%) |
| --- | --- | --- | --- |
| C K | 56.15 | 55.30 | 54.86 |
| N K | 12.83 | 12.92 | 13.99 |
| O K | 23.14 | 23.92 | 23.5 |
| Na K | 0.93 | 1.00 | 0.82 |
| P K | 3.17 | 3.02 | 3.00 |
| S K | 2.92 | 2.93 | 3.00 |
| Cl K | 0.87 | 0.90 | 0.83 |
| K K | 0.02 | 0.04 | 0.06 |

**Fig.5:** Purity analysis. (a) row 1: SDS-PAGE gel electrophoresis with Coomassie blue stain results before and after preferential crystallization of full capsids of Sf9 produced rAAV5 (left image), rAAV9 (middle image), and rAAV8 (right image), and preferential crystallization of empty capsids of Sf9 produced rAAV5 (left image), rAAV9 (middle image), and rAAV8 (right image) showing the capsid protein components, VP1, VP2, and VP3, and protein impurities before and after the preferential crystallization. row 2: the corresponding Western blot results for starting sample identifying only the viral protein components, VP1, VP2, and VP3 and confirming that the other protein bands as observed in SDS-PAGE electrophoresis as protein impurities from cell lysis. (See Supplementary **Fig. 19** for SDS-PAGE electrophoresis and Western blot results for Sf9 produced rAAV5 capsids with transgene length 4.68 kbp).

Experimental conditions for preferential crystallization of full capsids: 3.5% PEG8000, 1.5 M NaCl at pH 5.7 for rAAV5 for a starting sample with 35% full capsids; 4% PEG8000, 1 M NaCl at pH 6.3 for rAAV9 for a starting sample with 24% full capsids; 2.3% PEG8000, 0.3 M NaCl at pH 6.1 for rAAV8 for a starting sample with 37% full capsids. Experimental conditions for preferential crystallization of empty capsids: 2% PEG8000, 0.3 M NaCl at pH 5.7 for rAAV5 for a starting sample with 27% empty capsids; 2.5% PEG8000, 0.6 M NaCl at pH 6.3 for rAAV9 for a starting sample with 40% empty capsids; 4.5% PEG8000, 1.7 M NaCl at pH 6.1 for rAAV8 for a starting sample with 37% empty capsids. (b) Left: SDS-PAGE gel electrophoresis with Coomassie blue stain results before and after preferential crystallization of full capsids of HEK293 produced rAAV5 with a starting sample of 27% full capsids and preferential crystallization of empty capsids of HEK293 produced rAAV5 with a starting sample of 25% empty capsids. Right: the corresponding Western blot results for starting samples. Experimental conditions for preferential crystallization of full capsids: 3% PEG8000, 0.8 M NaCl at pH 5.7 for a starting sample with 30% full capsids. Experimental conditions for preferential crystallization of empty capsids: 1% PEG8000, 0.8 M NaCl at pH 6.3 for a starting sample with 23% empty capsids. (c) **Table 3** showing the average values of compositions (Supplementary **Table 2**) as obtained from EDAX analysis in three different locations on crystals of full capsids of rAVV5, rAAV9, and rAAV8 as shown in the corresponding SEM images of crystals of rAAV5 (left), rAAV9 (right), and rAAV8 (bottom) at higher magnification. Preferential crystallization conditions for full capsids: 3% PEG8000, 1.4 M NaCl, pH 5.7 for rAAV5 for a starting sample with 26% full capsids; 4% PEG8000, 1 M NaCl, pH 5.7 for rAA9 for a starting sample with 23% full capsids; 2.5% PEG8000, 0.3 M NaCl and pH 6.1 for rAAV8 for a starting sample with 35% full capsids.

Point EDAX (energy dispersive X-ray analysis) was performed to establish the elemental composition of crystals at different locations (**Fig. 5c**) and the corresponding average elemental composition is shown in **Table 3** (Supplementary **Fig. 20** and Supplementary **Table 2**). EDAX indicates that the major element in the crystal is carbon (atomic% ~ 55) followed by oxygen (atomic% ~ 23) and nitrogen (atomic% ~ 13). This results in a ratio of ~ 4.2 : 1.8 : 1 for C : O : N in capsid. This could not be compared with the literature as there is no report on elemental composition of capsid. This is consistent as carbon, and hydrogen are the two major constituent elements in any protein, followed by oxygen, nitrogen, and sulfur. Due to the low mass, EDAX does not detect hydrogen. The presence of a very small fraction of Na, K, and Cl (atomic% < 1) in crystals indicates that the crystal is high purity. Thus, preferential crystallization removes the salt impurities too along with protein impurities, and empty and partially filled capsids.

The crystallization process in hanging-drop may need 1 to 2 weeks for crystal to nucleate followed by 2–3 days for crystal growth, due to slow evaporation. In industrial crystallization, seed-crystals are used to induce/promote secondary nucleation, accelerating the time for crystal nucleation. Seeding experiments (**Fig. 6a**) were performed to determine the time required for the seed-crystals/new crystals to grow to a separable/filterable size (minimum ~ 10 µm).[56] **Fig. 6a** shows that the crystals take almost 1–1.5 h to grow by 50 µm. Thus seed-crystals will reduce the overall crystallization time from 2–3 weeks to < 2–3 h. This suggests that the full capsids can be selectively crystallized from the cell lysate in just 2-3 h irrespective of the volume of the cell lysate or length scale of the crystallizer. Whereas the conventional purification methods need much longer time to separate full capsids from the cell lysate material and are dependent on the volume of the column/centrifuge tube and the cell lysate. E.g., to process 1 L of cell lysate material, affinity chromatography needs > 45 h, cation exchange chromatography needs CEX- > 15 h, anion-exchange chromatography needs AEX- > 20 h, CEX-AEX needs > 30 h, CEX-AEX-SEC (size-exclusion chromatography) needs > 30 h, and UC-based purification needs > 30 h.[54,57–60] Additionally, AEX, CEX-AEX, and CEX-AEX-SEC need PEG precipitated/dialyzed/affinity chromatography purified material as a feed, which further increases its overall processing time, has poor yield/recovery. Likewise, UC needs PEG precipitated material as a feed, has poor yield/recovery, and is not scalable. In contrary, preferential crystallization has no such limitation and needs only partially purified material as feed as the crystallization process itself removes impurities, making it a suitable potential alternative.

At the end, yield of the crystallization process (i.e., percentage of the initial capsids that formed crystals) was calculated for different experimental conditions (Supplementary **Fig. 21**). The corresponding average yield (**Fig. 6b**) suggests that the yield is > 88% for all pH and serotypes and is weakly dependent on the serotypes. Whereas for other purification processes, the yield/recovery strongly changes with the serotypes. The yield of preferential crystallization process is much higher than that obtained in single-cycle anion-exchange chromatography (AEX- 50-70%), cation-exchange chromatography (CEX- ~ 80%) and UC-based purification (often << 50%) and is comparable to the steric-exclusion chromatography

(95%) and affinity chromatography (~ 95%).[15,17,54,57,61,62]

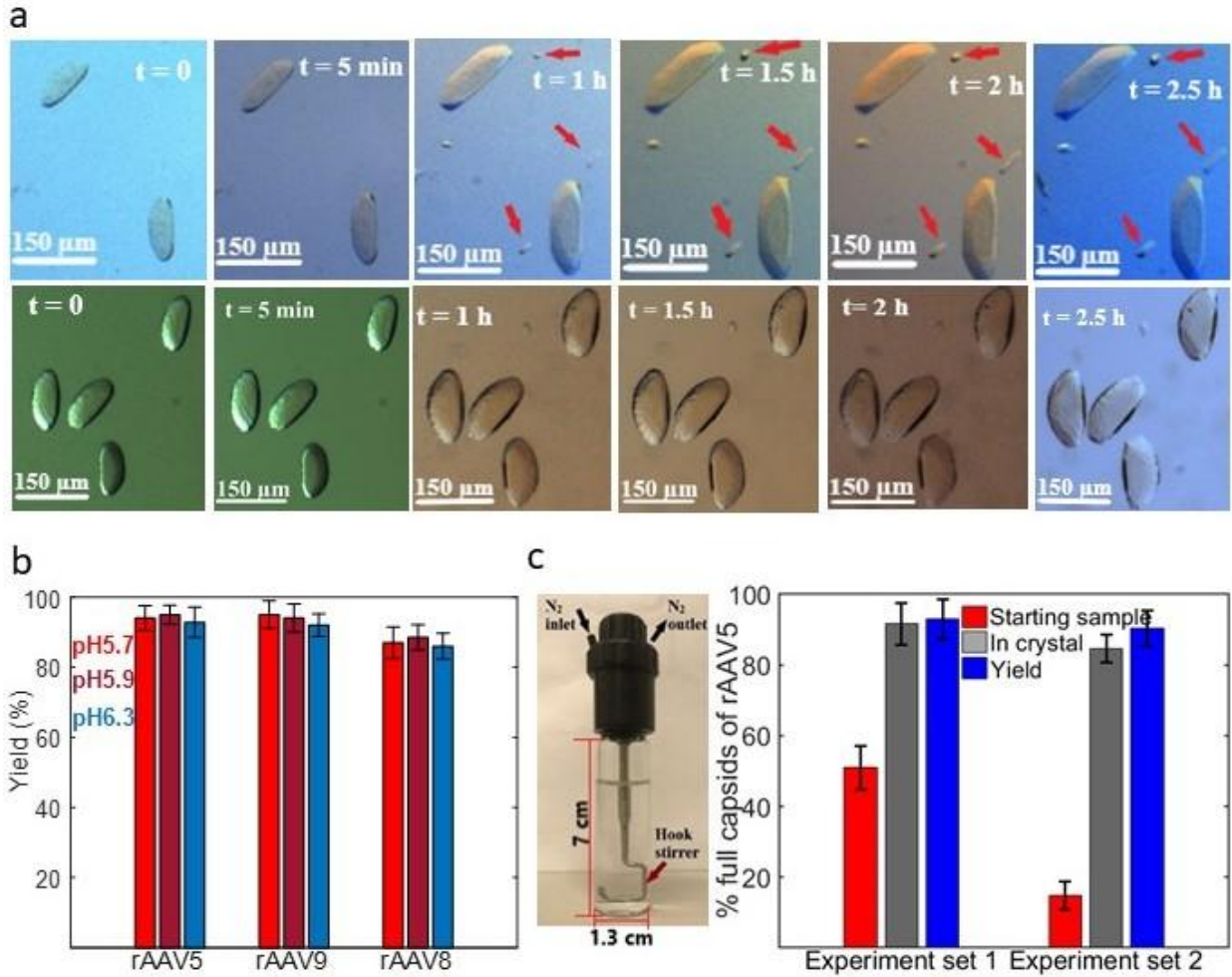

**Fig. 6:** (a) Growth of seed crystal and primary nuclei of full capsids of Sf9 produced rAAV5 (row 1) and rAAV9 (row 2) in the preferential crystallization region at different times. Experimental condition: 1.4M NaCl, 4% PEG8000 at pH 5.7 (row 1, rAAV5), and 1M NaCl and 3.75% PEG8000 at pH 6.3 (row 2, rAAV9). (b) Average of yields of preferential crystallization experiments performed at different crystallization conditions (as shown in Supplementary **Fig. 21**) in the preferential crystallization region of full capsids of Sf9 produced rAAV5 (left), rAAV9 (middle), and rAAV8 (right). Experimental conditions: Supplementary **Fig. 21**. (c) Preferential crystallization experiment in scaled up reactor (left) and the corresponding experimental result (right) for preferential crystallization of full capsids of Sf9 produced rAAV5. Experiment condition: 1.2M NaCl, 3.5%PEG8000 at pH 5.7; Concentration of rAAV5: $5 \times 10^{12}$ vg/mL ($2.5 \times 10^{13}$ capsids or 0.25 mg capsid /batch; set 1) and $1 \times 10^{13}$ vg/mL ($7 \times 10^{13}$ capsids or 0.7 mg capsid /batch; set 2). Batch volume: 5 mL (set 1) and 7 mL (set2)

To determine whether preferential crystallization works in an automated scaled-up crystallization device, experiments were performed in a 7-mL "crystallizer" (Crystalline, Technobis NL). **Fig. 6c** shows the crystallizer and corresponding experimental results for rAAV5 for starting sample with 51%, and ~ 15% full capsids. Results show that both the percentage of full capsids in the crystals and the yield are close to that found in hanging-drop experiment (**Table 1 Fig. 3a**). These scale-up results suggest that the preferential crystallization observed in µL-volume droplet experiments are reproducible in a larger-scale mL-volume reactor and supports the conclusion that the preferential crystallization is scalable. The conditions used for preferential crystallization experiment in a droplet or the 7-mL crystallizer can be used to perform preferential crystallization in crystallizers of any scale without further modification, as the thermodynamics of crystallization remain the same across scales. Both the full capsid enrichment and yield achieved in preferential crystal-lization is greater than that reported for one round of conventional CEX-AEX (< 75% full as observed in electron microscope image, 50–70% yield) and CEX-AEX-SEC (< 75% full as observed in electron microscope mage, < 50% yield).[54,58–60,63] AEX alone is only capable of a 50–70% yield with < 65% full capsids from a starting sample with almost 15% full capsids.[15,54,64,65] Similarly, CEX alone is only capable of almost 80% yield with poor full capsids enrichment for a starting sample with low full capsids.[54,60] UC may approach approximately 90% full capsid purity, but the recovery is often << 50%.[17,62] Sequential isopycnic gradients is used to achieve higher full capsid enrichment, but at a substantial loss of recovered vector. In addition, co-purifying impurities remain, which require additional process steps, e.g., SEC, and dialysis that further reduce yield. Scale-up results suggest that, starting with a low percentage of full capsids, two rounds of preferential crystallization can give a purity > 95% with a recovery (yield) almost 80% (**Fig. 6c**). In contrast, two rounds of AEX would give almost 50% recovery and two rounds of UC would give a recovery of << 25%.[17,62,64,65] Thus, the preferential crystallization method has the potential to be used in the downstream purification for removal of protein and salt impurities and enrichment of full rAAV capsids from cell lysate containing full, partially filled, and empty capsids, and other protein impurities, in a single step, in one round, and in a short period of time with full capsid enrichment, purify, and yield higher than the existing methods without the need of additional purification steps.

**METHODS**

Methods, including statements of data availability and any associated references, are available in the online version of the paper.

**SUPPLEMENTARY INFORMATION**

Any supplementary information and source data files are available in the online version.

## ONLINE METHODS

### Materials

All chemicals used in the experiments were of molecular biology grade. Chemicals were purchased from Sigma-Aldrich, sodium dihydrogen phosphate dihydrate (BioUltra, ≥ 99.0%), sodium hydroxide (BioXtra, ≥ 98% anhydrous), 1M hydrochloric acid (BioReagent, for cell culture), sodium chloride (BioXtra, ≥ 99.5%), potassium chloride (BioXtra, ≥ 99.5%), potassium dihydrogen phosphate (BioUltra, ≥ 99.5%), magnesium chloride (BioReagent, ≥ 97%), phosphate-buffered saline (PBS 1X (150 mM sodium phosphate and 150 mM NaCl), pH 7.2; BioUltra solution), polyethylene glycol (PEG-8000, BioUltra; PEG-6000, BioUltra), Phosphotungstic acid (10% solution), and Propidium iodide solution (1 mg/mL in water). Dulbecco's 1X PBS, FreeStyle 293 F cells, FreeStyle 293 expression medium, DMEM 1X (Dulbecco's modified Eagle medium), 0.25% Trypsin-EDTA (1X), FBS (fetal bovine serum) were purchased from Thermo Fisher Scientific. Nuclease-free water (Ambion nuclease-free water), Trypan blue stain (0.4%), and countess II automated counting slides were purchased from Invitrogen. Poloxamer-188 (Pluronic F-68, 10%, BioReagent) was added to rAAV-containing solutions at 0.001% to suppress binding to container surfaces. Glutaraldehyde solution (10% aqueous solution) and copper grid (carbon support film, 200 mesh, Cu; CF200-CU-50) were purchased from Electron Microscopy Sciences. Amicon Ultra centrifugal filter (10 kDa, UFC5010) and Ultrafree MC VV centrifugal filters (PVDF 0.1 µm; UFC30VV25) were purchased from Millipore Sigma. Falcon tube (5 mL, polystyrene round-bottom tube with cell-strainer cap), cell culture dish (10 cm × 2 cm), Erlenmeyer shake flask (125 mL), Costar 96-well cell culture plate (TC treated, sterile), and 6-well cell culture plate (TC treated, sterile) were purchased from Corning. An ELISA kit was purchased from Progen and stored in a 4°C refrigerator. DNA primers were purchased from Integrated DNA Technologies (Coralville, IA, USA). Eva Green Supermix was purchased from Bio-Rad Laboratories. The cover glasses and the silicon gaskets containing six sample wells for mass photometry measurement were provided by Refeyn Ltd. (MA, USA).

### rAAV samples

Recombinant adeno-associated virus serotypes, rAAV5 and rAAV9, both as full (genome-loaded) and empty (without genome) capsids, were purchased from Virovek at a reported concentration of $10^{14}$ vg/ml. The sample labeled as "full rAAV5" (Lot 19-047E) from Virovek actually contains 80% full and 20% empty capsids, and the sample labeled as "empty rAAV5" (Lot 19-253E) from Virovek contains 92% empty and 8% full capsids.[1] Similarly, the sample labeled as "full rAAV9" (Lot 21-100) from Virovek contains 80% full and 20% empty capsids, and the sample labeled as "empty rAAV9" (Lot 21-077) contains 96% empty and 4% full capsids. Virovek quantified the full and empty capsids using qPCR and nanodrop OD (optical density) measurement. In-house quantification of full and empty capsids using ddPCR and ELISA produced roughly similar results as reported by Virovek. The full rAAV capsids contained a genome of length 2.5kbp with cytomegalovirus (CMV) promoter expressing green fluorescent protein (GFP). For each type of sample, Virovek supplied the sample from the same batch, so the rAAV samples in our experiments do not have the variability that can occur when samples are taken from different batches. rAAV samples were supplied at a pH of 7.2 in PBS buffer in small vials holding 100 µL samples each, and were stored long-term at −80°C. Before putting the samples into the freezer, 100 μL sample from each vial was divided into 5 equal aliquots. Aliquots are stored at −80°C for long-term use. For short-term use, based on requirement, small vials were taken out of the −80°C freezer and stored at 4°C, at which AAV is stable for 4 weeks.[66,67]

---
[1] Reported by Virovek.

## Experiment
### Hanging-drop vapor-diffusion experiment

Crystallization conditions for rAAVs are screened in a hanging-drop vapor-diffusion experiment (Supplementary **Fig. 3**) using VDX 24-well crystallization plates with glass cover slips on top (Hampton Research, California, USA). Each well contains two liquid solutions: one being a droplet of small volume of 2 μL suspended from a glass cover slip at the top of the well, and the other being a reservoir of 1 mL at the bottom of the well. Each drop and reservoir contain polyethylene glycol (PEG) 8000 and sodium chloride (NaCl) as precipitants dissolved in a phosphate-buffered saline (PBS) solution. The reservoir is to ensure that the liquid conditions in each droplet would reach quasi-steady conditions sometime after closing the system (Supplementary **Fig. 3**).

Each droplet was a mixture of 1 μL of AAV sample and 1 μL of buffer solution. Initial screenings were performed by varying the concentration of PEG8000 (from 0.1 to 7 w/v%) with a constant NaCl concentration. Subsequent screenings were conducted by varying the NaCl concentration to fine-tune the ionic strength of the crystallization medium. As time progresses, water evaporates from the droplet. Sometime after supersaturation is reached, the nucleation of capsid crystals begins and eventually a vapor-liquid equilibrium is achieved at which point no further reduction in the volume of the droplet takes place.

Each droplet was monitored at regular intervals for a period of 1 to 2 weeks via an optical microscope (Imaging Source DMK42BUC03) to track the evolution of crystals. The particles are observed in real time using a microscope with an in-built CCD camera (Leica Z16 APO), using both normal and polarized light. Capsid nucleation took 1-2 weeks to complete, which was followed by growth of the crystals for up to 2 to 3 days; at which point no further change in the crystals was observed. All experiments are performed for pH between pH 5.5 to 8.5 where the rAAV particles are stable (low degradation).[67,68] All experiments were performed at room temperature, 23±2°C.

### Construction of phase diagram

To find crystallization conditions (i.e., the concentration of precipitants PEG and NaCl at a specific pH), where only either full or empty capsids are crystallized, >500 crystallization conditions are screened for "full" capsids and "empty" capsids samples each for specified values of the pH. At each tested condition, the AAV capsids either remain in solution, form an amorphous precipitate, or form crystals. A phase diagram is then constructed/formed/prepared by plotting these data as a function of PEG and NaCl concentrations. Separate phase diagrams are determined for "full" and "empty" capsids samples at different pH values. To screen 500 crystallization conditions, the whole parameter space is discretized into different regions where each region represents an experimental condition. This approach allows exploring one parameter at a time keeping other parameters constant. Based on the outcome of an experiment, the next experiment is performed at higher concentration (higher spacing) or lower concentration (lower spacing). Please visit the Supplementary **Fig. 4** for exemplary phase diagrams and corresponding experimental points.

### Crystallinity confirmation

X-ray diffraction (XRD) is the most commonly used method to confirm the crystallinity of a material.[69–72] Single crystal X-ray diffraction was performed with crystals looped out from the droplet, but because of the extremely high molecular weight and resulting dense structure, it was not successful. Because XRD requires a relatively large sample mass (~ 0.2 g), for limited samples, crystallinity is determined by two different alternative methods that require much less sample mass: (i) light polarization, and (ii) scanning electron microscopy (SEM) imaging of crystals.

**Polarization study.** The optical properties of the crystals were assessed using normal and cross-polarized light microscopy. Screening of crystallization conditions based on birefringence in cross-polarized light microscope is much simpler and faster than destructive methods such as SEM and TEM, making the approach suitable for repetitive screening of a large number of conditions. Crystals appear as colorless or grey in normal light microscopy images (Supplementary **Fig. 7a**). A 24-well plate was positioned on a Leica Z16 APO cross-polarized light microscope base and each droplet was observed. Then plane-polarized light is passed through the crystals and the analyzer, which is also a polarizer, is rotated 360° while keeping the crystal's position fixed. On rotation of the analyzer, anisotropic crystals will assume a spectrum of interference color except at every 90° position of the analyzer, where the interference is extinct and the crystals appear dark/extinct (Supplementary **Fig. 8**). This phenomenon is known as bire-

fringence and confirms the crystallinity of particles.[47-50]

When crystals are observed under cross-polarized light, crystals remain dark/extinct (Supplementary **Fig. 8**) at every 90° position (where two polarizers' vibration directions are perpendicular to each other) and show interference coloration on rotation due to the splitting of the transmitted light into two rays: slow (e) and fast moving (o) rays.[73–75] These photos (Supplementary **Fig. 8**) indicate that the crystals are optically aniso-tropic (i.e., the speed of light varies with direction) and birefringent (i.e., depends on the polarization and propagation direction of light). [73,75] The presence of a unique angle of extinction and birefringence in the crystals in this analysis suggests that the particles are single crystals, in contrast to polycrystalline materials which would appear isotropic.[73,76] Under cross-polar-ized light, isotropic crystals do not show any birefrin-gence as light passes through without splitting, and the crystals appear dark (extinct), because of the absence of interference.[73] Similarly, materials that are not crystals (e.g., protein aggregates) do not show any birefringence.[73] The literature suggests that NaCl generally forms cubic crystals (isotropic) and does not show any interference colors under cross-polarized light microscope.[73,75] Thus, rAAV crystals can be easily distinguished from NaCl crystals if NaCl is present in the system. All AAV capsid crystal particles generated in this study are optically anisotropic, and are not cubic in shape, which indicate that the particles are not NaCl crystals.

**SEM imaging of AAV crystals.** For SEM analysis of AAV crystals, conditions that produce well-developed crystals with sharp edges are considered. The droplet is diluted with equal volumes (5 µL each) of reservoir solution and 1X PBS buffer solution (pH 7.2) to prevent the dissolution of crystals and facilitate manipulations. The solution is agitated carefully to detach the crystals from glass surface without breaking the crystal. Then the crystals and solution are aspirated and dispensed on a centrifugal filter of 0.1 µm pore size (Millipore Sigma) and centrifuged once at 2000g for 1 min to remove the free capsids and solution. Then the membrane filter paper is cut out of the tube frame and attached on a carbon tape on a brass stub. Then a 10-nm thick gold coating (EMS Quorum, EMS 150T ES, MIT Material Science) is applied on the crystal sample to make the surface conducting and then taken for SEM analysis. SEM analysis is performed in a high-resolution electron microscope (ZEISS merlin, MIT Material Science).

**Crystal collection, dissolution and stock solution preparation for biological analysis (ELISA, ddPCR, cell infection, flow cytometry, SDS PAGE, mass photometry)**
Stock solution of rAAV is prepared by dissolving crystals for biological assays. To prepare AAV stock solution, crystals are collected from the droplet following the same procedure as described in the sample preparation for SEM imaging and dispensed into a centrifugal filter paper of 0.1 µm pore size and centrifuged at 2000g for 1 min at ambient temperature. The crystals remain attached to the filter paper, effect-tively removing free AAV particles, PEG, and NaCl that remain in the supernatant. The crystals are washed twice (100 µL of 1:1 mixture of reservoir solution and 1X PBS) to ensure the complete removal of residual virus capsids and to dilute the PEG and NaCl. The number of washing steps may vary based on the type of crystals in the droplet as AAV crystals are very soft and prone to breakage even with low shear. For small crystals, 1/2 washing steps were performed to avoid breakage and eventual passage of crystals through the filter membrane pores. Then the crystals are washed once with 100 µL of diH$_2$O (≥18 mΩ, MilliQ) and centrifuged at 2000g for 1 min to remove the adsorbed PEG and NaCl from the filter. The centrifugation is performed immediately after addition of diH$_2$O as crystals are highly soluble in water. The AAV crystals are then dissolved in 300 µL of nuclease-free water 94°C refrigerator for 30 min). The dissolved crystal solution is transferred into an Amicon centrifugal filter (10 kDa NMWCO) and centrifuged at 10000g for 5 min. Capsids are then washed 3 times using 200 µL of nuclease-free water at 10000g for 5 min each time to ensure complete removal of remaining salt and precipitant. The capsid particles are then resuspended into 200 µL of Dulbecco's 1X PBS (DPBS without magnesium and calcium) and collected into a vial and stored at 4°C for short-term use. For long-term stor-age, 5% sorbitol solution is added to the capsid sus-pension and stored in a cryovial at –80°C.[77]

**Calculation of percentage of full and empty capsids in crystals**
**ELISA experiment.** ELISA assay protocols followed the manufacturer's instructions (Progen) for rAAV5 (AAV5 Xpress ELISA kit) and AAV9 (AAV9 Xpress ELISA kit). Briefly, to determine a workable capsid concentration for ELISA assays, a range of concen-trations is produced by serial two-fold dilution of the AAV stock solution (see section for stock solution preparation): 1X, 2X, 4X, 8X, 16X, 32X, 64X, 128X, 256X, 512X, 1024X. Capsid concentrations outside the workable range, will either lead to signal saturation or low signal when the concentration is too high or too low, respectively. Assay buffer (ASSB 20X, Progen) is diluted to ASSB 1X using diH$_2$O (milli-Q). Similarly, (lyophilized) capsid standards, known as

'kit control' (KC), are included in the ELISA kits. The KC is reconstituted using ASSB 1X (500 µL) and a serial dilution (1X, 2X, 4X, 8X, 16X, 32X, 64X) is prepared using ASSB 1X. Anti-AAV5 mAb-biotin conjugate is dissolved in ASSB 1X (750 µL) to produce 20X Biotin solution. For both KC dilution and sample dilution, duplicates are prepared. KC dilution (100 µL) and AAV stock dilution (100 µL) are loaded on ELISA plate and the assay is performed following the procedure as recommended by the ELISA kit manufacturer. At the end of the assay, absorbance of the solution in each well is read immediately in a microplate reader (Synergy H1; BioTek Instruments, Winooski, VT, USA) at 450 nm wavelength. Absorbance readings for the sample dilution as well as KC dilution are corrected by subtracting the absorbance of the control. The averaged absorbance values, from duplicates, for the KC dilution (along the y-axis) are plotted against the corresponding concentration along the x-axis and a 4-parameter logistic (4PL) fit is performed in Matlab to calculate the concentration of the capsids in the sample.

**ddPCR assay.** The quantity (i.e., titer) of genomes was determined using ddPCR assays. Like ELISA, a serial dilution (1X, 2X, 4X, 8X, 16X, 32X, 64X, 128X, 256X, 512X, 1024X) of AAV stock solution (see Section for stock solution preparation) was performed using DNase-free water. A master mix solution (18.2 µL for each well) is prepared by mixing 7.79 µL of 2.62% DMSO, 0.41 µL of primer mix (forward and reverse), and 10 µL of EvaGreen Supermix (QX200 ddPCR EvaGreen Supermix; BIO-RAD). In each well of the 96-well PCR plate, diluted AAV stock solution (2.2 µL) and the master mix (18.2 µL) are combined. Standard thermal cycles for the EvaGreen supermix set by BIO-RAD: 1 cycle of enzyme activation step performed at (95°C for 5 min). 40 cycles of amplification cycle consisting of denaturation (95°C for 30 s), annealing and extension (60°C for 1 min). Following the amplification cycles, a signal stabilization cycle, is performed (4°C for 5 min for 1 cycle and then at 90°C for 5 min). Finally, the sample is held at 4°C before being removed from cycler. Forward primer was

$$5' - GCAAAGACCCCAACGAGAAG - 3'$$

and reverse primer was $5' - TCACGAACTCCAGCAGGACC - 3'$.

DNase-free water is used as the non-template control and for each dilution, duplicate sample is prepared. The number of viral DNA copies in the stock solution and in the crystals are back calculated from the ddPCR data.

**Room temperature TEM analysis of AAV particles after crystallization**

To visualize and measure the percentage of full and empty capsids in the crystals as well as to understand whether the virus particle's morphology remains the same, room temperature TEM imaging was performed. Crystals are collected following the procedure described in the sample collection for SEM imaging and washed following the procedure as described in stock solution preparation section. The AAV crystals are dissolved by adding diH$_2$O (100 µL) onto the filter (30 min at room temperature). The solubilized capsids are fixed by adding 2.5% aqueous glutaraldehyde solution in 1:1000 v/v ratio to inactivate the capsids.

Virus particles images are obtained by TEM (JEOL model) at room temperature (TEM facility, Koch Institute, MIT). For sample preparation, 10 µL of dissolved crystal solution is aspirated and dropped on a 200-mesh copper grid coated with continuous carbon film and at room temperature. After 10 minutes, any remaining water on the grid is absorbed by a tissue paper (Kimwipes) and 10 µL of 1% aqueous phosphotungstic acid solution is immediately dropped onto the moist TEM grid. After 30 s, the excess liquid is absorbed by a tissue paper. The grid is left at room temperature for 45 minutes to allow the remaining liquid to dry before inserting the grid into the TEM chamber. All images are recorded on a Gatan 2k×2k UltraScan CCD camera. Empty capsids show the contrast difference as shown in the Supplementary **Figs. 11a-l**. The heavy metal stain is taken up by the empty capsids which diffracts the electron beam appearing as black areas (Supplementary **Fig. 11**, black arrows) in the image whereas full capsids exclude the heavy metal atoms and appear as uniformly shaded hexagons (Supplementary **Fig. 11**, white arrows).

**Mass photometry experiment.** Mass photometry measurements are performed on a SamuxMP instrument (Refeyn Ltd., MA, USA), which has a higher resolution tailored especially for AAV vector analysis. Prior to each measurement, a calibration is performed using "empty" AAV9 vector (3.74 MDa). The molar mass of the "empty" AAV9 is provided by Refeyn Ltd. (MA, USA). To generate the calibration curve, 10 µL of 1X PBS is pipetted into a well of silicon gasket, the focus is adjusted automatically, and then 10 µL of AAV9 calibrant is added into the loaded PBS. Before measurement, the loaded sample and PBS are mixed vigorously by aspirating and dispensing multiple times carefully to avoid formation of bubbles. The measurement time is set to 60 s, which captures the binding and unbinding events in the form of a movie. The measurements are recorded using Acquire MP 2.4.2 (Refeyn Ltd., MA, USA) and analyzed with DiscoverMP (v2023 R1.2) (Refeyn Ltd., MA, USA). For all the measurements, the binning width is set to 40, and the ratiometric contrast distribution is fitted by a Gaussian function to obtain the molecular weight of subpopulation. The F/E ratio (full capsid to empty capsid ratio) and linearity are visualized using MATLAB.

### Capsid protein sequencing and PTMS (post-translational modification) analysis
**LC-MS.**
For mass spectrometry analysis, protein is first digested overnight, passed through a chromatography column, and analyzed in orbitrap mass spectrometer (Thermo Fisher Scientific). For protein digestion, 8M urea, 50 mM Tris-HCL, 15 mM iodoacetamide, and 1% formic acid solutions are prepared and rAAV capsid protein sample from Virovek is digested using Trypsin Gold, Mass spectrometry grade following the procedure as suggested by Promega. Digested protein is then analyzed in Mass spectrometer (Orbitrap) following the protocol as suggested by manufacturer (Thermo Fisher Scientific).

### Crystal purity analysis
**SDS-PAGE electrophoresis analysis of capsid proteins.** SDS-PAGE analysis is performed to assess the quality of the AAV samples and determine whether crystallization affects the viral proteins integrity. Viral proteins are denatured and reduced by incubating a mixture of AAV sample, 2-mercaptoethanol (along with SDS, dye and glycerol)), and diH$_2$O (2:1:2 volume ratio) at 95°C for 10 minutes. The denatured samples and the protein ladder (BenchMarkTM, Thermofisher) are resolved electrophoretically through a 4–20% polyacrylamide gel (BIO RAD TGX gradient gel in a Mini-PROTEAN tetra electrophoresis cell/chamber) in 1X Tris/glycine/SDS buffer at 100 V for 100 minutes. For silver staining, fixative enhancer solution, development accelerator solution, silver staining solution (7:1:1:1 mixture of diH$_2$O, silver complex solution, reduction moderator solution, and image development solution), and stop solution (5% acetic acid solution) are prepared and gel is treated as per manufacturer's instruction (BIO RAD), except the final staining step, where gel is stained for 7–10 minutes. The silver-stained gel is imaged in ChemiDoc imaging system (BIO RAD). For Coomassie blue staining, the gel is stained as per the manufacturer's instruction (Invitrogen) and imaged in ChemiDoc imaging system.

**Western blot.** Denatured rAAV sample and the protein ladder are resolved electrophoretically following the procedure as mentioned for SDS-PAGE electrophoresis. Protein is transferred (blotted) from gel to polyvinylidene fluoride (PVDF) membrane using wet transfer method in a tank in 1X Tris-glycine buffer following the procedure as described by ABCAM. For antibody application, Anti-AAV VP1/VP2/VP3 Mouse monoclonal, B1 primary antibody (Progen Biotech), Rabbit Anti-Mouse IgG H&L (HRP) secondary antibody (ABCAM), SuperSignal Western Blot Enhancer (Invitrogen) are purchased. Then blocking buffer (5% nonfat dry milk in 1X TBST and 2.5% nonfat dry milk in 1X TBST) is prepared and the membrane is treated as per the instruction of ABCAM. After primary and secondary antibody application, membrane is imaged in ChemiDoc imaging system.

**EDAX (energy dispersive X-ray analysis) analysis.** EDAX analysis of crystals is performed (EDAX, AMETEK materials analysis division) using an Octane Elect Super Detector attached with SEM microscope (ZEISS Merlin). Sample preparation remains the same as for SEM imaging of crystals except there is no gold coating on crystals for EDAX analysis. For EDAX analysis, a crystal is first selected and then an elemental mapping is done on the area, which includes both crystal and the underlying membrane surface. Elemental mapping is performed on the membrane surface as a control. Finally, line and point EDAX is performed on different locations on a crystal. For all the analysis, more than 100 scans are used unless mentioned. Crystals analyzed in EDAX are more than 10 µm thick and a beam of 15 kV is used for this analysis.

**Cell biological activity measurement. Preparation of adherent cell culture.** Transduction in cell culture is used to determine whether crystallization affects biological activity of rAAV. Cryopreserved HEK293T (ATCC) cells are thawed (2 min at 37°C water bath) and transferred into a centrifuge tube that contains 9 mL of complete growth medium (10 v/v% FBS in DMEM). The cells are then pelleted (centrifuged at 1000g for 5 minutes) and resuspended in complete growth medium three times to remove the DMSO used in cryopreservation. The cell pellet is then resuspended in the preconditioned complete growth medium (9 mL at 37°C in CO$_2$ for 15 min), dispensed in a 10-cm cell culture dish (Corning tissue culture-treated culture dishes), and then incubated at 37°C in 5% CO$_2$ atmosphere (passage 1). The culture is observed under microscope twice daily.

Adherent cells, 80% confluent, are detached from the cell culture dish using trypsin / EDTA solution (3 mL of 0.25% Trypsin-0.53 mM EDTA) for 10 min and transferred into centrifuge tubes after addition of complete growth medium (10 mL). Cells are then pelleted (1000g for 5 min) and resuspended in complete growth medium (3 mL) thrice to remove trypsin. Finally, cell pellet is resuspended into 1 mL of complete growth medium and viable cell density is measured in a Countess II cell counter (Invitrogen, MA, USA) with trypan blue (0.4% stock) 1:1 ratio.

A subculture is prepared following the same procedure as described for passage 1 and the remaining cell suspension is cryopreserved in 10% DMSO solution for future use.

**Study of HEK293T cell transduction by rAAV present in crystals.** For study of cell transduction, cells from passage 1 (2 mL in each well of a Corning Costar TC-treated 6-well plate) are transduced with AAV stock solution (see section for stock solution preparation) for multiplicity of infection (MOI, number of viral genome-containing particles per cell) of $10^4$, $10^5$, and $10^6$. Cells are then cultured in an incubator at 37°C in 5% $CO_2$ atmosphere and observed in a fluorescence microscope once daily for 5 days. On day 5, viable cell density and percentage of GFP-positive cells are measured in a Countess II Cell Counter with trypan blue (1:1; v:v, ratio of 0.4% stock). Transduction efficiency, as a percentage, is calculated by dividing GFP-positive cells by the total cell number × 100%.

**TCID50 experiment.** Transduction experiments, analogous to tissue culture infectious dose (TCID50) virus assays are performed to obtain a more precise value of biologically active vector titer. In this experiment, ~5000 cells (HEK293T cells from passage 1 as described in adherent cell culture preparation section) are seeded in each well of a 96-well plate (Corning Costar TC treated) and cultured in complete growth medium (100 µL 10%FBS in DMEM) for 2 days in a humidified incubator at 37°C and 5% $CO_2$. On day2, a dilution series of rAAV stock solution (see stock solution preparation section) is prepared using complete growth medium in a 96-well plate (Corning Costar V shaped well) with dilutions of $10^{-3}$, $10^{-4}$, $10^{-5}$, $10^{-6}$, $10^{-7}$, $10^{-8}$, $10^{-9}$, and $10^{-10}$. 100 µl of each dilution is added to each well in a column (8 wells) except for the first two columns, which are used as a negative control. The cells are then cultured (at 37°C and 5% $CO_2$) and cell growth and GFP fluorescence are examined in a microscope. On day 5, the percentage of GFP-positive wells are measured and used to calculate the "TCID50" according to the Reed-Muench algorithm.[78]

**Suspension cell culture preparation.** Flow cytometry is performed to obtain a more accurate measurement of cell viability and percentage of GFP-positive cells than microscopy-based approaches. Cryopreserved HEK293 cells are thawed at room temperature and dispensed into 500 µL pre-warmed (at 37°C and 5% $CO_2$) FreeStyle 293 expression medium and cultured in a Corning Erlenmeyer cell culture flask in 20-mL FreeStyle 293 expression medium. Cell viability and cell density are measured once daily in a Countess II Automated Cell Counter (Invitrogen) using trypan blue (0.4% stock) as previously described. Cells are passaged when viable cell density reaches 2 to 3 million per mL.

**Flow cytometry experiment.** For flow cytometry experiments, a total of nine subcultures are prepared by adding 1.1 mL HEK293 cells (from passage 21 with 99% viability; ~ 0.25 million viable cells) in 50 mL FreeStyle 293 medium (at 37°C and 5% $CO_2$) in each Erlenmeyer shake flask (Sigma Aldrich). Out of the nine subcultures, one shake flask serves as a negative control and four shake flasks serves as a positive control, where cells are transduced with standard AAV sample (Virovek) at the following ratios of particles to cells or MOIs (multiplicity of infection): $10^6$, $5\times10^5$, $10^5$, and $10^4$. In the remaining four shake flasks, cells are transduced with the AAV stock solution (see the section for stock solution preparation) at the same MOIs as in the positive control. Cells are incubated in an orbital shaker (at 37°C, 5% $CO_2$, and 135 rpm) and aliquots collected at 24 h intervals for flow cytometry experiments. Aggregates and cell debris are removed from the samples (0.5 mL for each experiment) with nylon mesh cell strainers in polystyrene tube. The cell viability and % GFP-positive cell numbers are measured in Aria 4 (Flow core, Koch Institute, MIT) using PI (propidium iodide) as a viability dye.

**Seeding experiments**

Nucleation (primary nucleation) followed by growth of crystals in hanging drop vapor diffusion experiments (in a 24-well plate) takes about 2 weeks to complete. For the crystallization method to be useful as an alternative purification method, the process needs to be completed within a time comparable to that of existing technologies. To evaluate the feasibility of crystallization for AAV purification, a crystal seed growth experiment is used to reduce the overall time by eliminating the requirements for nucleation. In a seeding experiment, purified/precleaned crystal seeds are added into the crystallization medium at a certain time. To mimic the seeding experiment, a droplet containing well-grown crystals was identified, and growth solution (2 µL of 1:1 ratio of reservoir solution and AAV sample) was added. The variation in size of crystals is measured at 5 min to 1 h intervals in a microscope (Leica Z16 APO), and the time required for the crystals to grow to a size (minimum 10–100 µm) for it to be separable in industrial filters was measured.

## Yield calculation

For the calculation of yield, droplets containing only a few distinct large crystals with no precipitate are selected and a small volume (5 µL) of reservoir solution is mixed thoroughly with the original droplet solution. Then 1 µL of the solution is collected, diluted 1000X by dispensing into PBS buffer, and analyzed to determine the quantities of total (ELISA) and full capsids (ddPCR). Subtracting full capsids from total capsids determines the empty capsid titer. The yield is the difference between the final and initial quantity of capsids divided by the initial quantity of capsids

## Scale-up experiment

To understand whether the preferential crystallization is relatively scalable, a microliter droplet-based system was scaled up to a milliliter-volume crystallizer device (Crystalline, Technobis, NL), which is widely used for small-molecule pharmaceuticals. Like the hanging drop system, the scale-up crystallizer also has an evaporation capability but, due to the flow of inert gas through the crystallization chamber, evaporation is faster in the latter. Unlike the hanging droplet system, the scaled-up reactor has a cooling jacket and a built-in stirrer. For this experiment, crystallization solution (5 mL 1:1 mixture of the rAAV sample and precipitant solution) is incubated at 30°C for 3 hr for 24 hr. Solution is simultaneously evaporated by passing $N_2$ gas throughout the chamber. A small volume (100 µL) of the solution is observed under microscope and ddPCR and ELISA experiments are performed.

## Capsid surface potential measurement

**Zeta potential experiment.** A series of solutions of different pH (as shown in Supplementary **Fig. 2**) for each ionic strengths 0.15 M and 0.2 M NaCl was prepared using diH$_2$O. 100 µL rAAV sample from the Virovek was centrifuged at 14000g for 5 minutes in an ultrafiltration tube to remove the PBS buffer. rAAV is then resuspended in 50 µL of diH$_2$0 of different pH and ionic strength. Each sample was loaded in a zeta potential measurement cell and the zeta potential was measured in DynaPro ZetaStar (Department of Chemistry, MIT) following the procedure as described by Wyatt Technology (MA). Measured zeta potential was plotted against the pH for each ionic strength to obtain the isoelectric point.

## ACKNOWLEDGEMENTS

Funding is acknowledged from the MLSC, Sanofi, Sartorius, Artemis, and USFDA (75F40121C00131). Andreas Lucas Gimpel is acknowledged for training the first author in the earlier crystallization experimental work. Moo Sun Hong is acknowledged for preliminary training in use of the polarizer instrument. Jose Sangerman is acknowledged for training in the first experiment of suspension cell culture preparation. John Joseph, Emmanuel Kagning Tsinda, and Jose Sangerman are acknowledged for training in the first ddPCR, and ELISA experiments. Help of Prasanna Srinivasan in the first adherent cell culture preparation is acknowledged. Help of Rui Wen Ou in the first electrophoresis experiment is acknowledged. Operators Young Zhang and Zhen-yuan Zhang at MIT.Nano and DongSoo Yun, and Daniel Cham Chin Lim at the Koch Institute are ac-knowledged for discussion of crystallography and material characterization methods.

## COMPETING INTERESTS

The authors have filed a patent related to this work.

## AUTHOR CONTRIBUTIONS

V.B. conceptualized this work, designed and conducted the experiments, analyzed the data, and wrote the initial draft. P.W.B. and R.D.B. conceptualized the work and contributed to designing experiments, supervising the work/project, editing the manuscript, and acquiring funds to support the project. S.L.S. contributed to supervision of the work/project, editing the manuscript, and acquiring funds to support the project. A.J.S. contributed to supervision of the work and acquiring the funds to support the project. R.K.M. helped design experiments, editing the manuscript, and acquiring the funds to support the project. J.M.W. contributed to supervising the work, editing the manuscript, and acquired the funds to support the project.

# Supporting information for Selective Enrichment of Full AAV Capsids

Vivekananda Bal[1], Jacqueline M. Wolfrum[2], Paul W. Barone[2],
Stacy L. Springs[2], Anthony J. Sinskey[2,3], Robert M. Kotin[2,4], and Richard D. Braatz[1,2]

[1] Department of Chemical Engineering, Massachusetts Institute of Technology, Cambridge, MA, USA

[2] Center for Biomedical Innovation, Massachusetts Institute of Technology, Cambridge, MA, USA

[3] Department of Biology, Massachusetts Institute of Technology, Cambridge, MA, USA

[4] Gene Therapy Center, University of Massachusetts Chan Medical School, Worcester, MA, USA

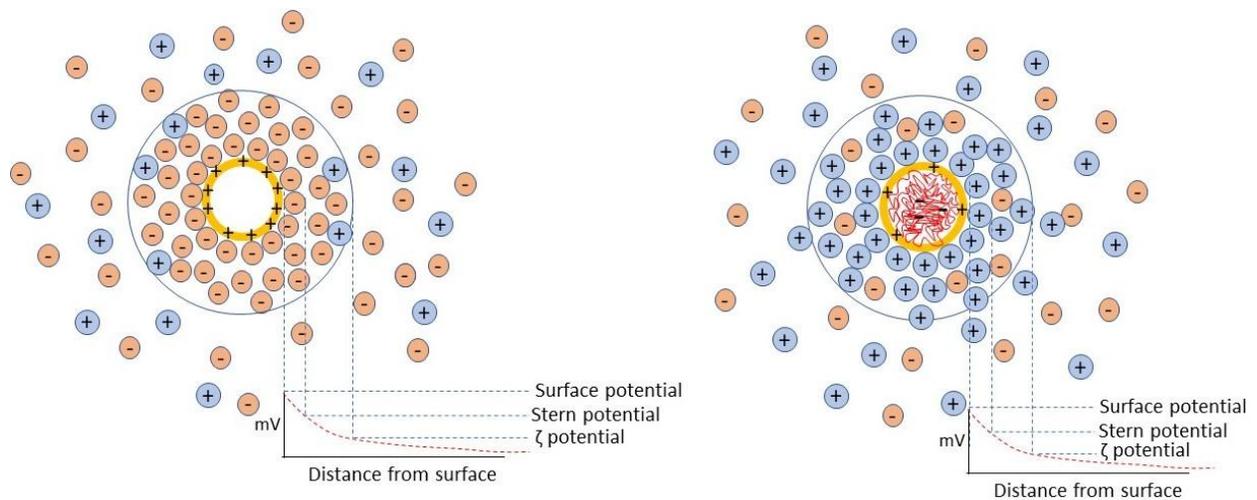

**Fig. 1:** A schematic of solution behavior of empty (left) and full (right) capsids. Capsids original charge density as determined by the surface amino acid composition and pH and the inside DNA load is surrounded by the oppositely charged salt ions. Formation of electrical double layer surrounding a capsid and screening of capsids original charge density, which results in the decay of influence of original surface charge and the decay of surface potential with the distance from the surface

*When a charged/colloidal particle is immersed in water, a layer of strongly bound oppositely charged ions surrounding the particle is formed and another layer of diffused loosely bound ions of similar and oppositely charged ions surrounding the first layer is formed. These two layers are collectively known as "electrical double layer." The potential at the outer surface/slipping plane of the diffuse layer is known as "zeta-potential."*

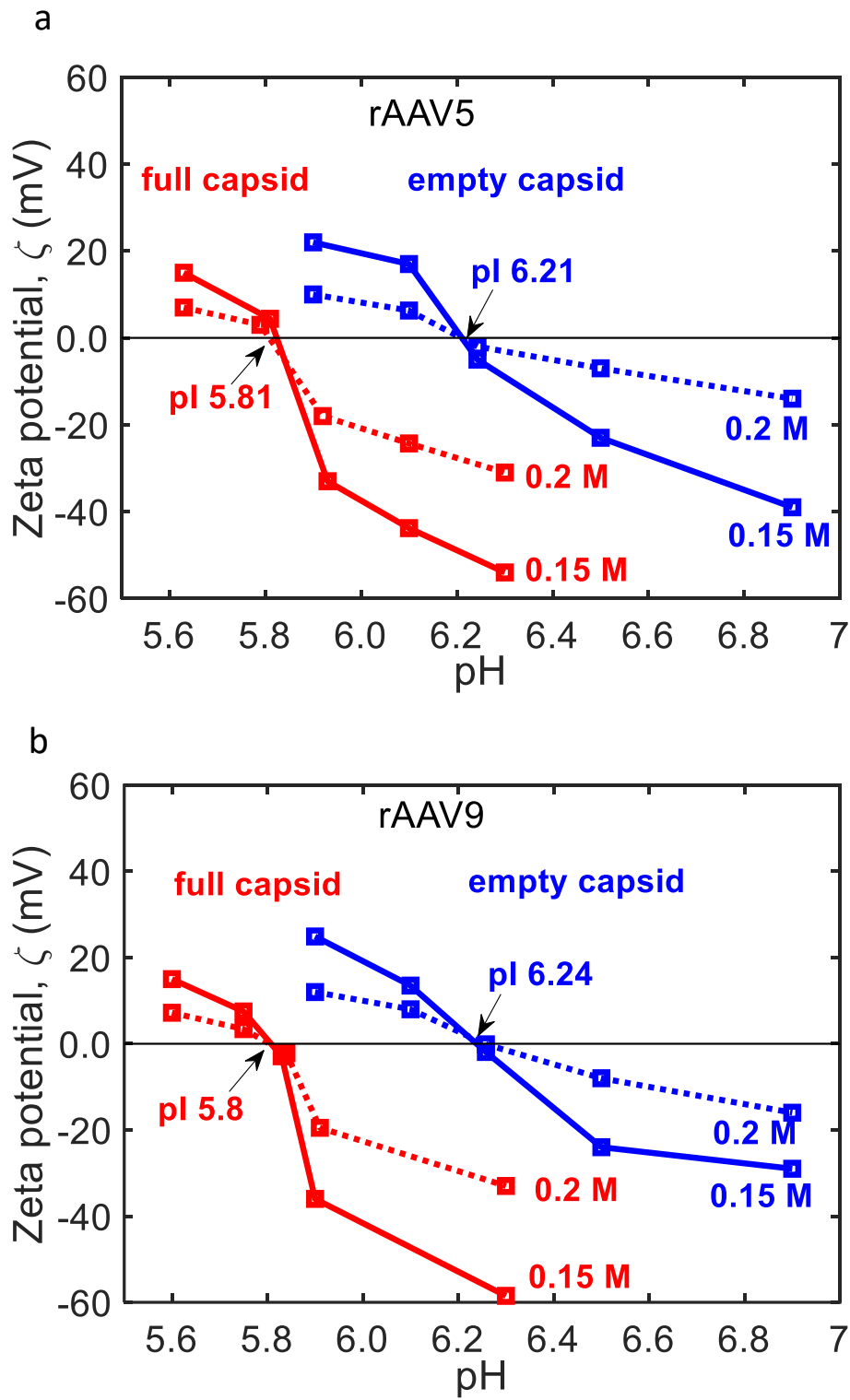

**Fig. 2:** Variation of zeta potential of full and empty capsids of rAAV5 (a) and (b) as a function of pH and ionic strength

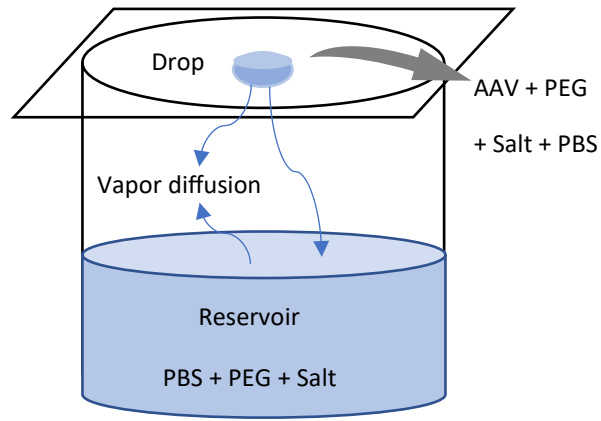
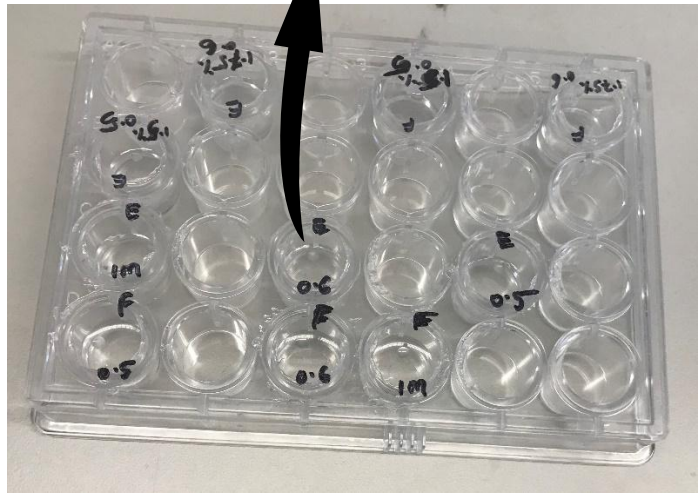

**Fig. 3:** Hanging-drop vapor diffusion experiment in 24-well crystallization plate

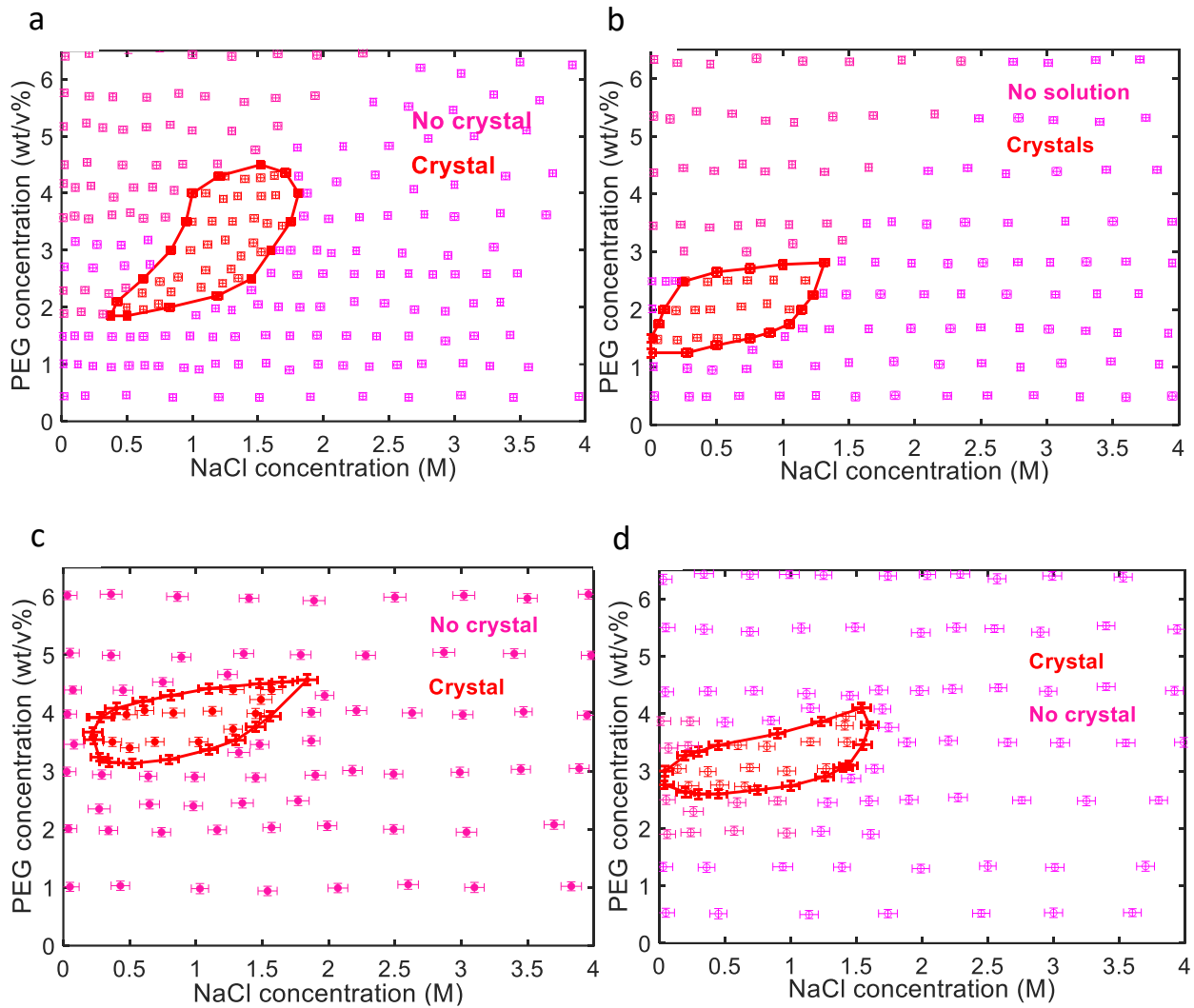

**Fig.4:** Phase diagram showing the regions favorable for crystallization, precipitate formation, and undersaturated solution as a function of PEG8000 and NaCl at pH 5.7 as an example for sf9 produced "full" rAAV5 sample (a), "empty" rAAV5 sample (b), "full" rAAV9 sample (c), and "empty" rAAV9 sample (d). Large number of screened crystallization conditions, which is essential for construction of phase diagram, are shown by experimental data points with error bars. Data points are showing the experimental error bars from three repeated runs.

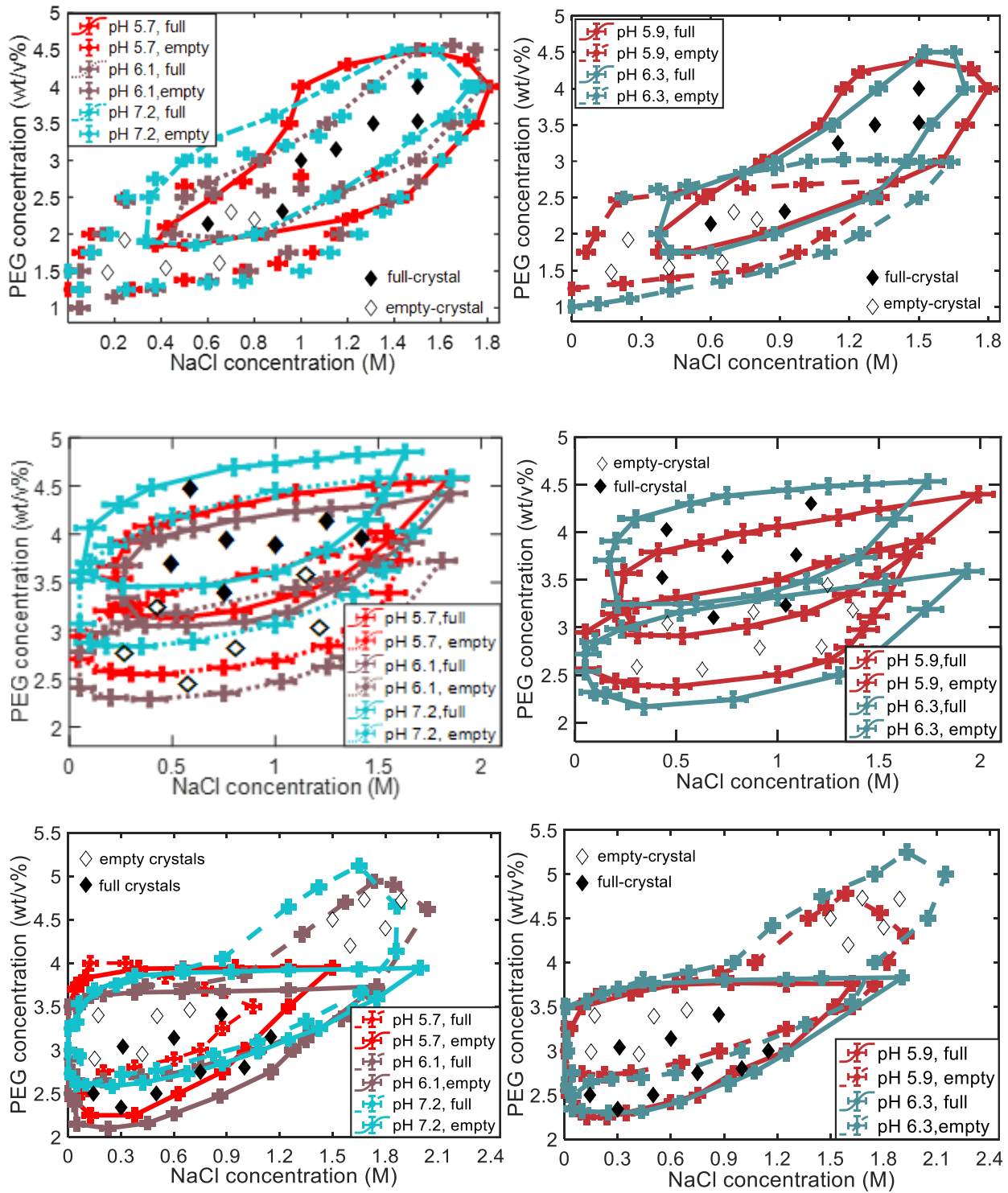

**Fig. 5**: Preferential crystallization region for full capsids and empty capsids of sf9 produced rAAV5 (top row), rAAV9 (middle row), and rAAV8 (bottom row) at different pH as a function of PEG8000 and NaCl concentrations

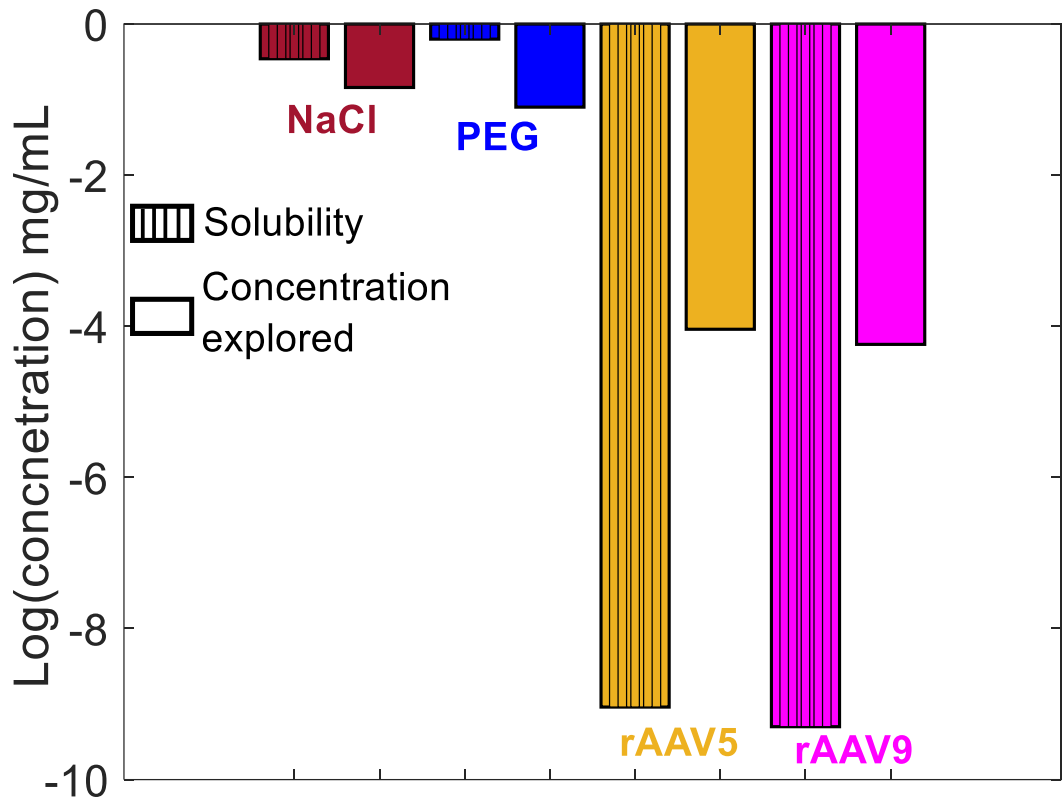

**Fig. 6:** Comparison of solubilities of NaCl, PEG8000, and rAAV with their concentrations used in crystallization process.

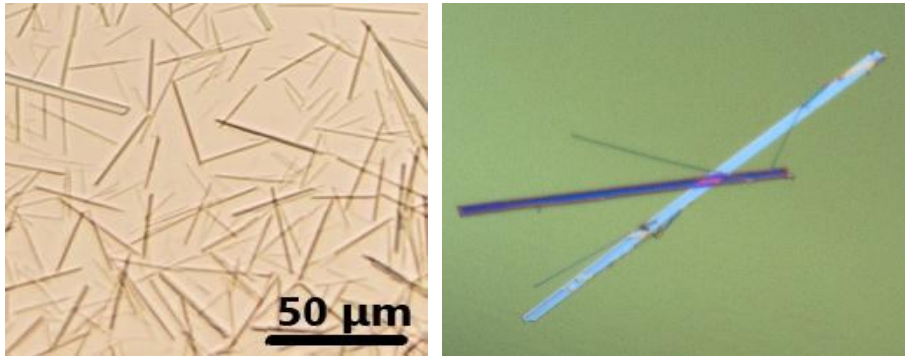

**Fig. 7:** Sample microscopic image of the droplet, showing, empty AAV5 crystals in (a) normal and (b) polarized light. Experimental condition: 1.5% PEG8000, 0.125 M NaCl, pH 5.7

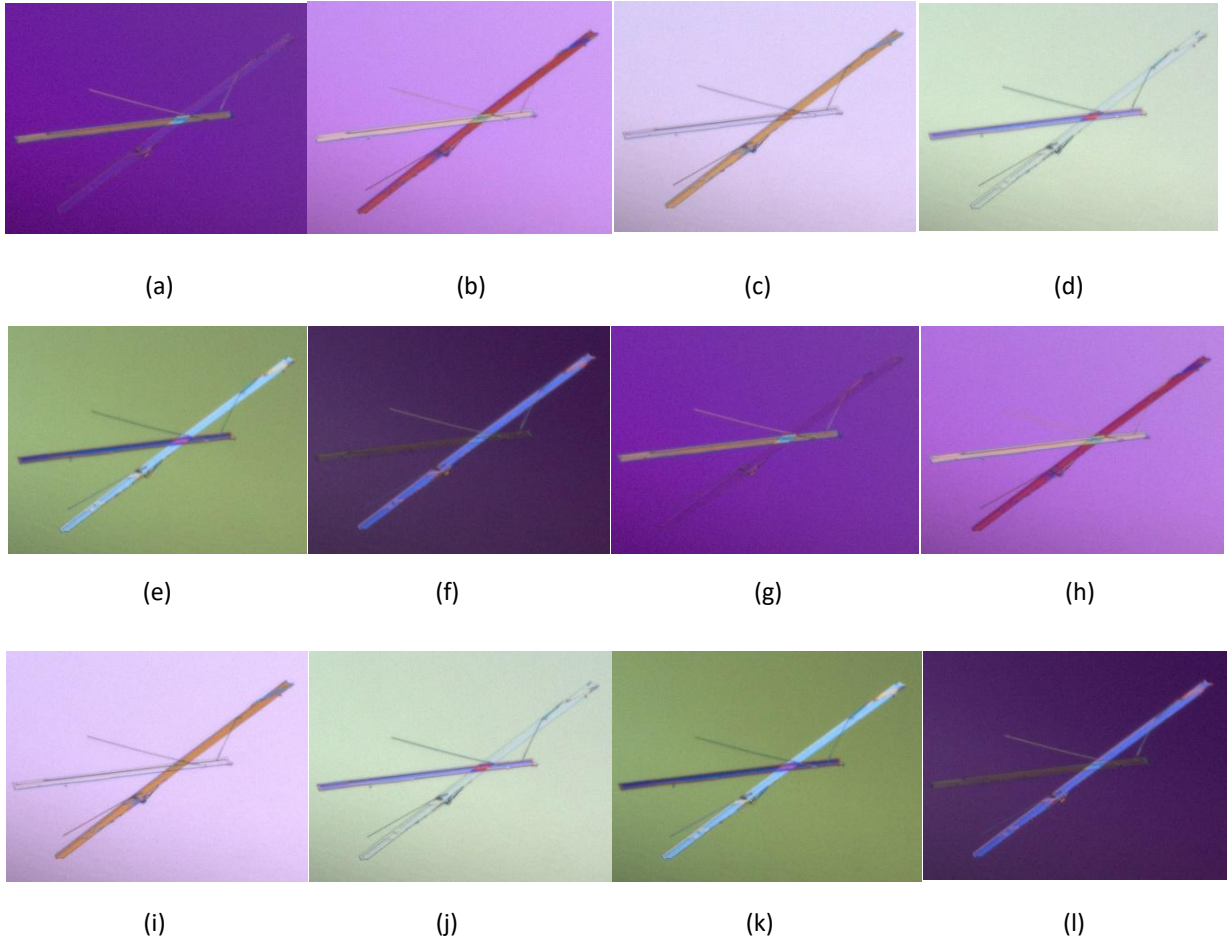

**Fig.8:** Images of crystals observed through cross-polarizer when crystal is fixed on the base of microscope and analyzer is rotated at an angle of (a) 0º, (b) 30º, (c) 60º, (d) 90º, (e) 120º, (f) 150º, (g) 180º, (h) 210º, (i) 240º, (j) 270º, (k) 300º, and (l) 330º

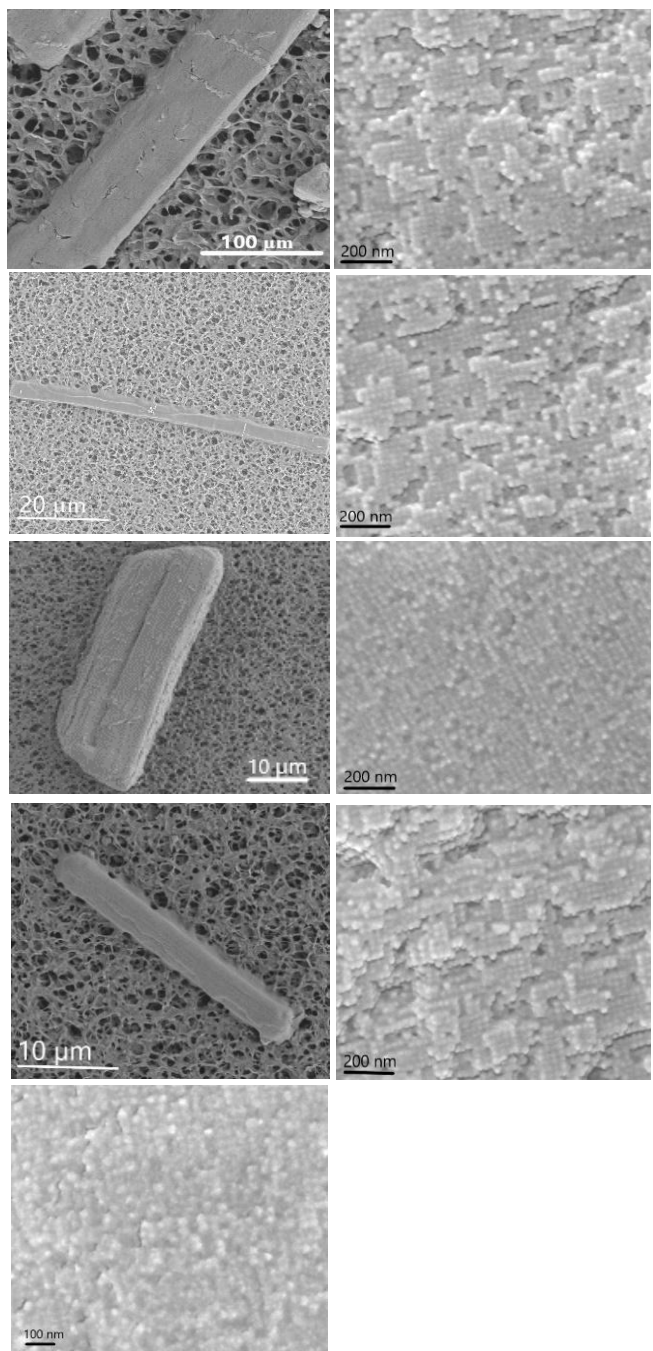

**Fig.9:** Row 1: SEM image of crystals of empty rAAV5 (left) and corresponding SEM image of crystal surface at higher magnification (right); 1.5% PEG8000, 0.4 M NaCl, pH 5.7 for a starting sample with 31% empty capsids. Row 2: SEM image of crystals of full rAAV5 (left) and corresponding SEM image of crystal surface at higher magnification (right); 3% PEG, 1.4 M NaCl, pH 5.7 for a starting sample with 26% full capsids. Row 3: SEM image of crystals of "empty" rAAV9 (left) and corresponding SEM image of crystal surface at higher magnification (right); 2.75% PEG, 1 M NaCl, pH 5.7 for a starting sample with 25% empty capsids. Row 4: SEM image of crystals of "full" rAAV9 (left) and corresponding SEM image of crystal surface at higher magnification (right); 4% PEG, 1 M NaCl, pH 5.7 for a starting sample with 23% full capsids. Row 5: SEM image of a precipitate. In all the experiments, length of the transgene was 2.46 kbp.

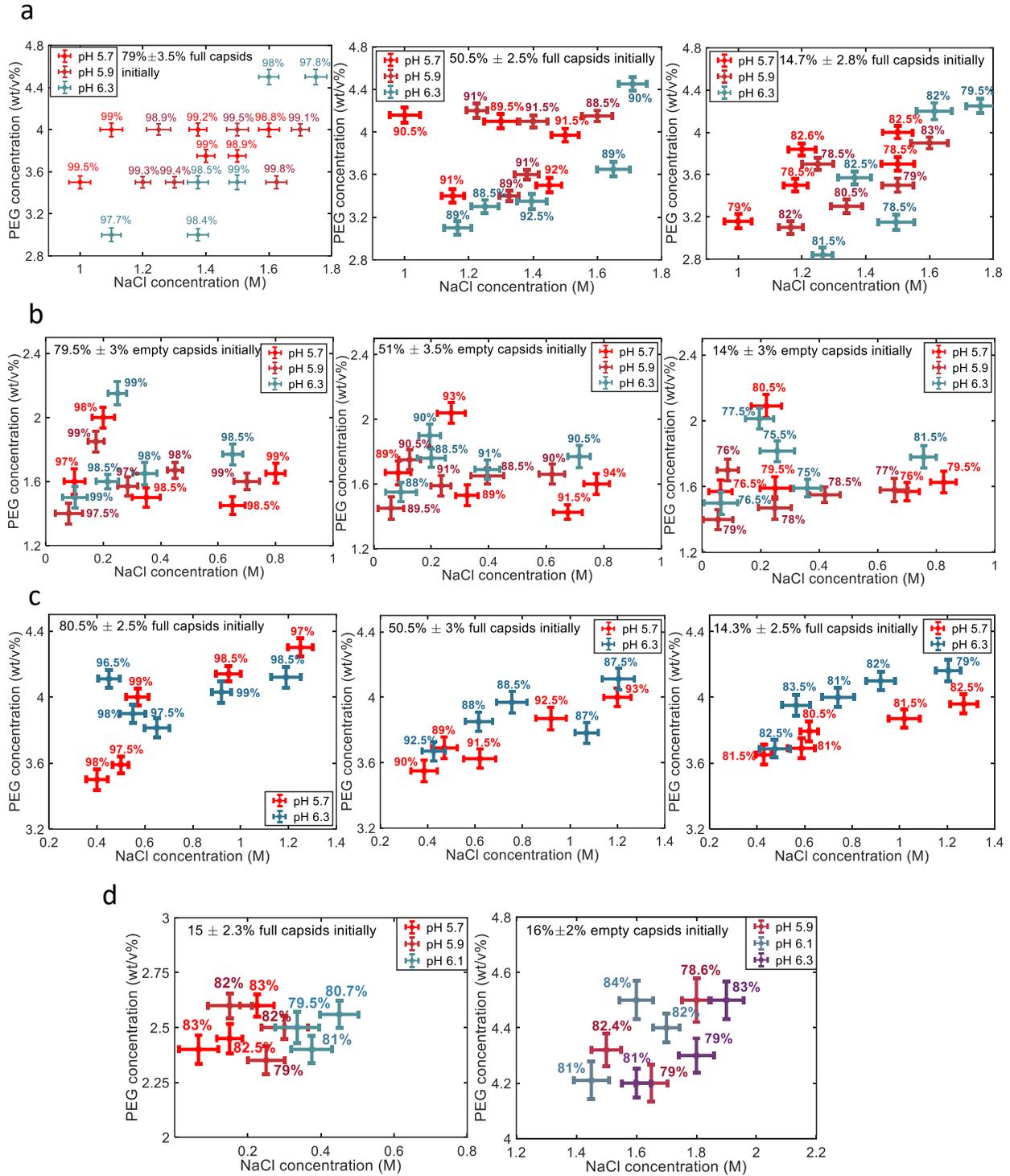

**Fig.10:** Percentage of full and empty capsids in the crystals obtained in preferential crystallization region as measured by ddPCR and capsid ELISA experiment. (a) Percentages of full capsids in the crystals obtained after crystallization in different conditions in the preferential crystallization region of full capsids of sf9 produced rAAV5 for starting sample with different initial concentration of full capsids. (b) Percentages of empty capsids in the crystals obtained after crystallization in different conditions in the preferential crystallization region of empty capsids of sf9 produced rAAV5 for starting samples with different initial concentration of empty capsids. (c) Percentages of full capsids in the crystals obtained after crystallization in different conditions in preferential crystallization region of full capsids of sf9 produced rAAV9 for starting samples with different initial concentration of full capsids. (d) Percentages of full capsids (left) and of empty capsids (right) in the crystals obtained after crystallization in preferential crystallization region of full capsids (left) and of empty capsids (right) of sf9 produced rAAV8. All experiments were performed with PEG8000 as a precipitant. In all the experiments, the length of the transgene was 2.46 kbp.

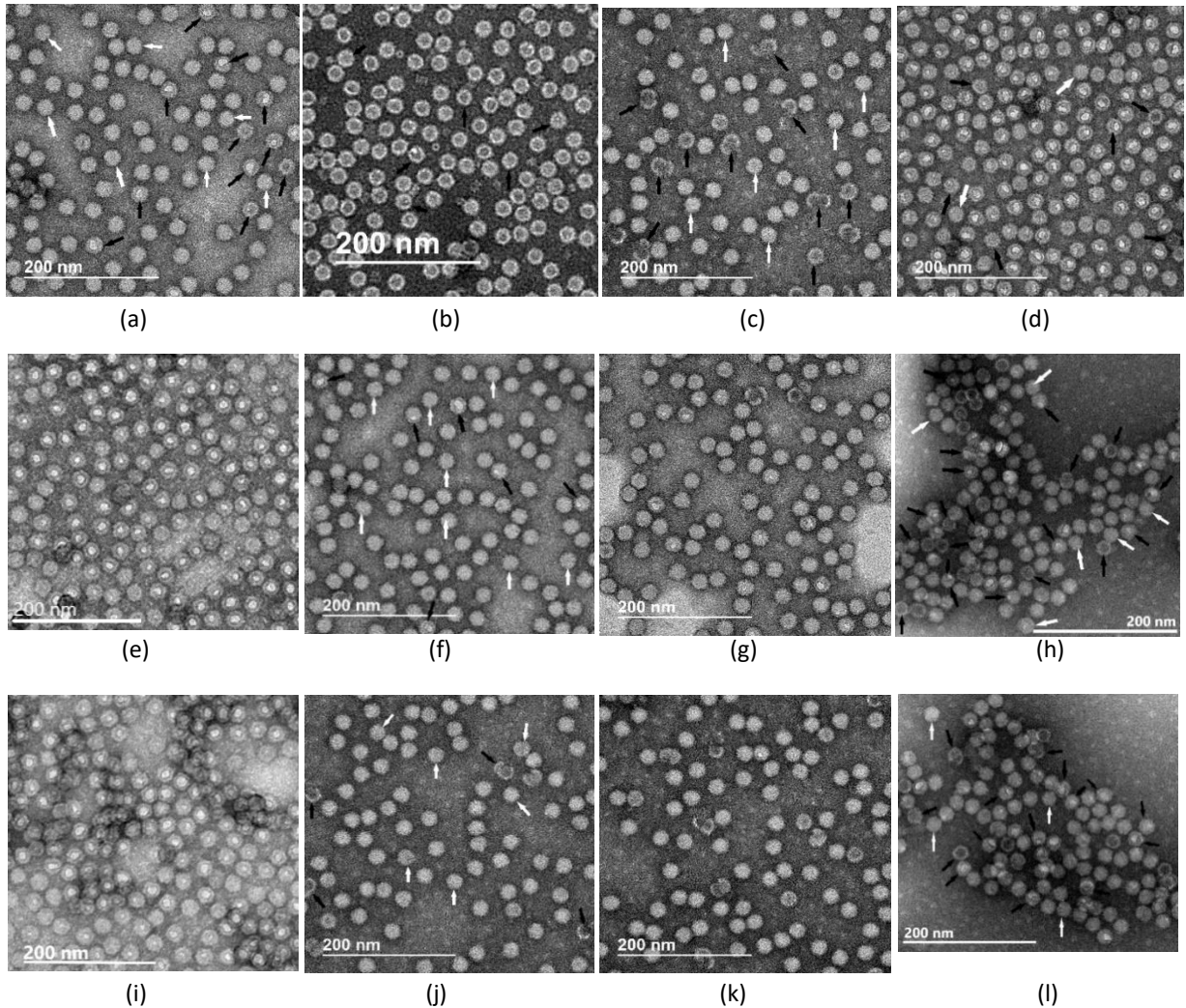

**Fig.11:** TEM image of negative stained sample of sf9 produced rAAV before and after preferential crystallization. Row1: (a) standard "full" rAAV5, (b) standard "empty" rAAV5, (c) standard "full" rAAV9, and (d) standard "empty" rAAV9; Row2: (e) rAAV5 sample with ~ 20% full capsids before crystallization, (f) stock solution of rAAV5 crystals obtained after full capsid's preferential crystallization from sample (e), (g) rAAV5 sample with ~ 20% empty capsids before crystallization, (h) stock solution of rAAV5 crystals obtained after empty capsid's preferential crystallization from sample (g); Row3: (i) rAAV9 sample with ~ 20% full capsids before crystallization, (j) stock solution of rAAV9 crystals obtained after full capsid's preferential crystallization from sample (i), (k) rAAV9 sample with ~ 20% empty capsids before crystallization, (l) stock solution of rAAV9 crystals obtained after empty capsid's preferential crystallization from sample (k). Experiment condition: Row2: 1.3 M NaCl, 3% PEG8000 at pH 5.7 for full capsids preferential crystallization (f), and 0.2 M NaCl, 2% PEG8000 at pH 5.7 for empty capsids preferential crystallization (h); Row3: 4% PEG8000, 1 M NaCl at pH 5.7 for full capsids preferential crystallization (j), and 2.75% PEG8000, 0.5 M NaCl at pH 5.7 for empty capsids preferential crystallization (l).

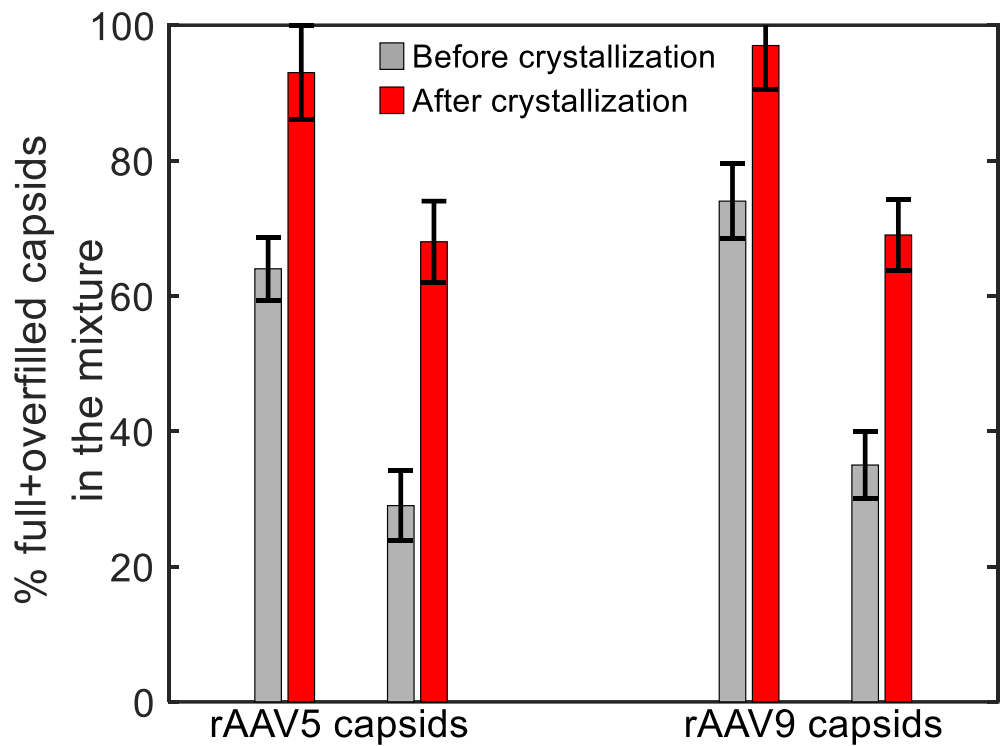

**Fig. 12:** Bar plots showing the percentage of full + overfilled capsids in the mixture before and after preferential crystallization of full capsids of sf9 produced rAAV5 and rAAV9 as obtained in Refeyn mass photometry experiment. Experiment condition: 4.2%PEG8000, 1.2M NaCl at pH 5.7 for rAAV5 and 3.75%PEG8000, 1.2M NaCl at 5.7 for rAAV9

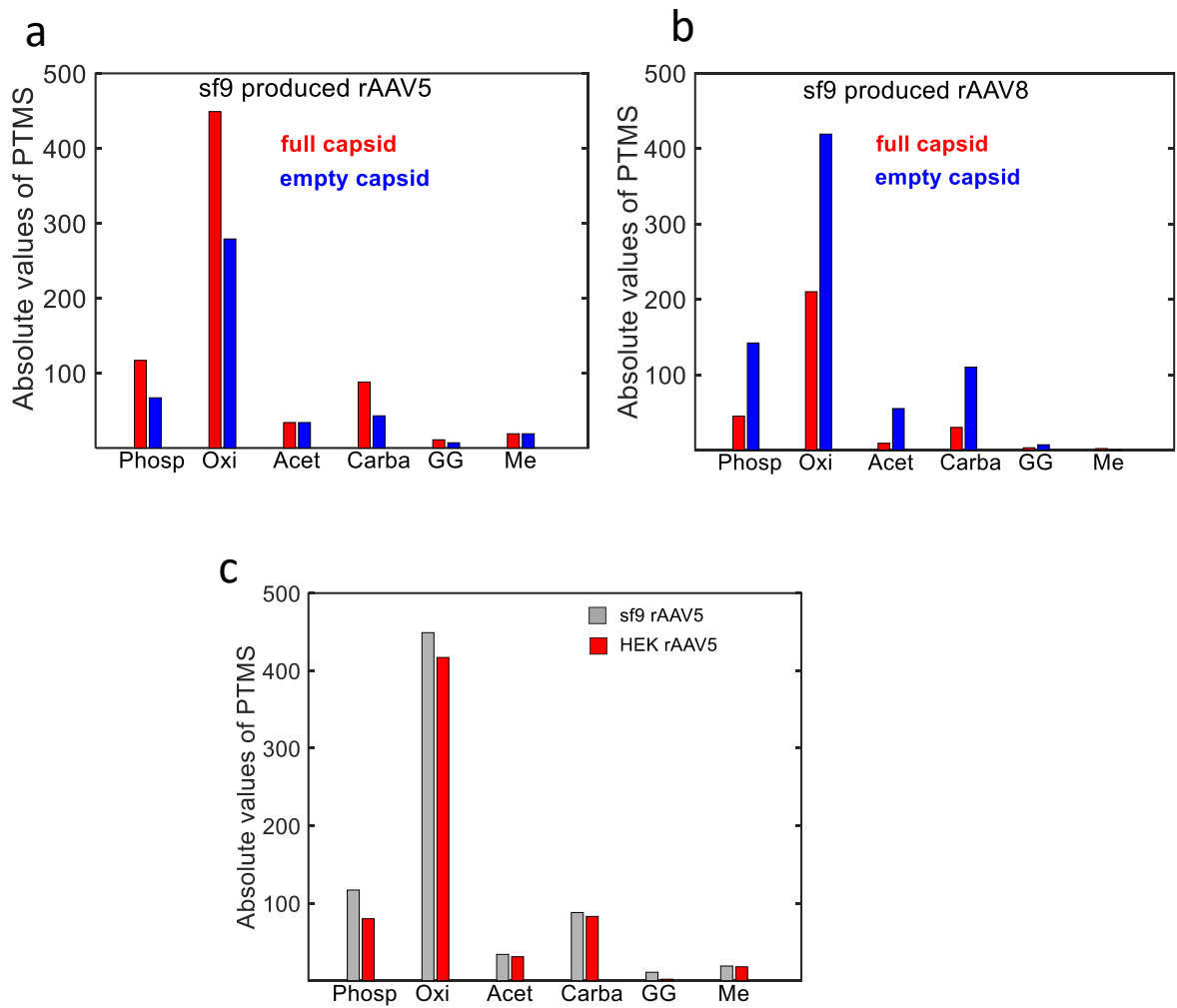

**Fig. 13:** Absolute values of post translational modifications (PTMS) in capsid proteins. (a) Comparison of PTMS values between full and empty capsids of sf9 produced rAAV5 capsids, (b) Comparison of PTMS values between sf9 and HEK produced rAAV8 capsids, (c) Comparison of PTMS values between sf9 and HEK produced rAAV5 capsids.

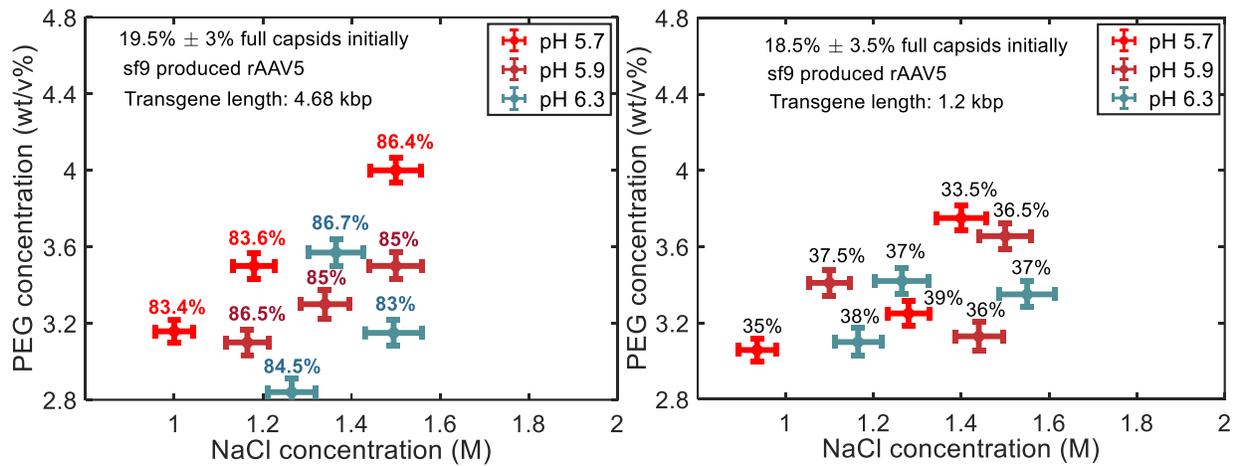

**Fig.14:** Percentage of full capsids in crystals obtained after crystallization in different experimental conditions in the preferential crystallization region for full capsids of sf9 produced rAAV5 of transgene length 4.68 kbp (left) and 1.2 kbp (right). PEG8000 was used as a precipitant.

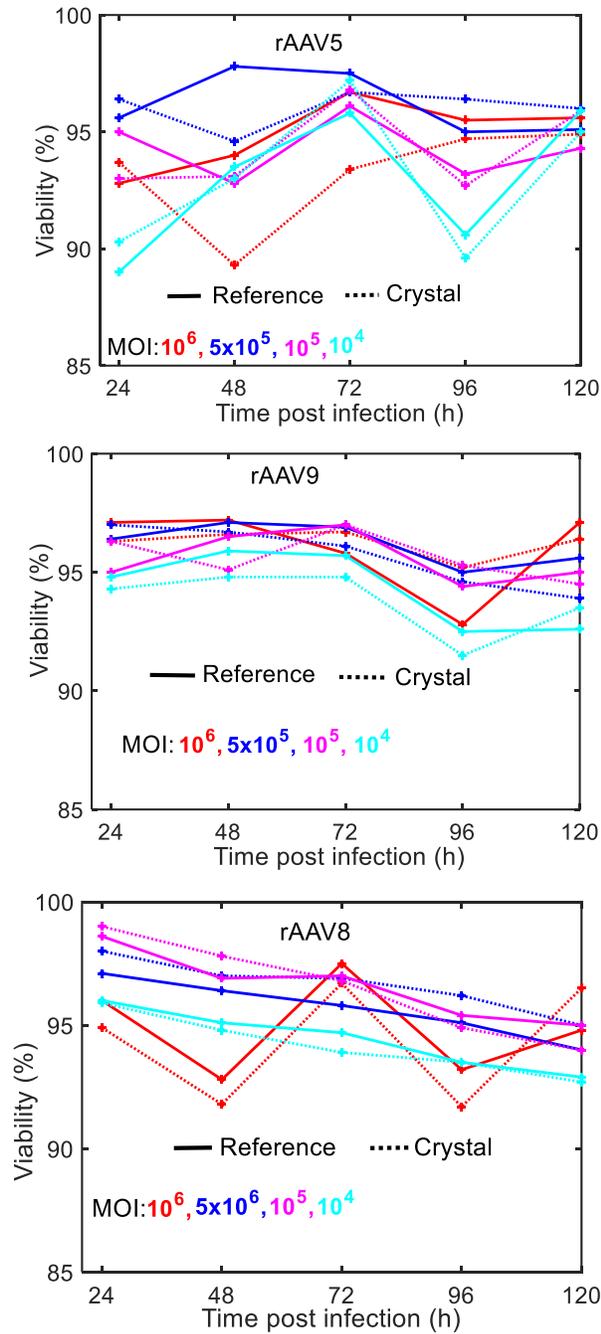

**Fig. 15: V**ariation of viability of HEK293 cells as a function of post transduction time for the transduction of HEK293 cell suspension with crystal stock solution as prepared from the crystals obtained after preferential crystallization of full capsids of sf9 produced rAAV5 (a), rAAV9 (b), and rAAV8 (c) and transduction with the reference sample as purchased from Virovek. For all the serotypes, full capsid's transgene length was 2.46 kbp. Experimental conditions for preferential crystallization of full capsids: 4% PEG8000, 1.2 M NaCl at pH 5.7 for rAAV5 for a starting sample with 33% full capsids; 4% PEG8000, 0.6 M NaCl at pH 6.3 for rAAV9 for a starting sample with 37% full capsids; and 2.5% PEG8000, 0.6 M NaCl at pH 6.1 for rAAV8 for a starting sample with 41% full capsids.

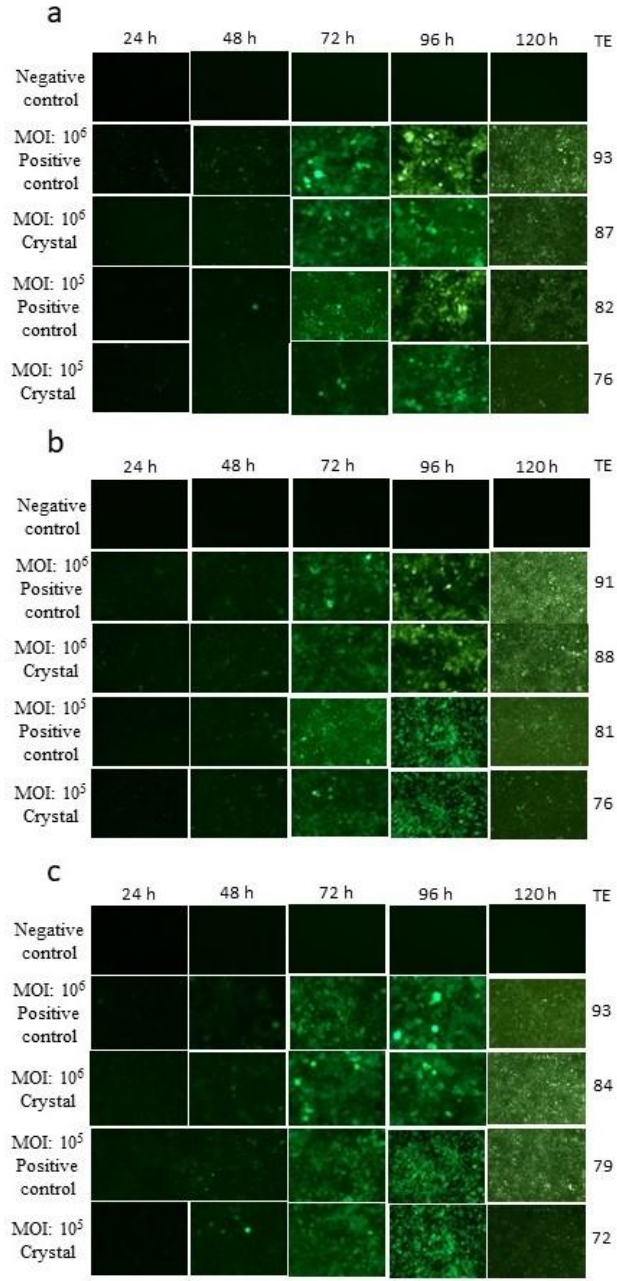

**Fig.16:** Fluorescence image for transduction of HEK293T adherent cells with full capsids of sf9 produced rAAV5 (a), rAAV9 (b), and rAAV8 (c) in crystal stock solution as prepared from the crystals obtained after crystallization in preferential crystallization region of full capsids. For all the serotypes, full capsid's transgene length was 2.46 kbp. Experimental conditions for preferential crystallization of full capsids: (a) 1.4M NaCl and 3.5% PEG8000 at pH5.9 for rAAV5 for a starting sample with 45% full capsids, (b) 1 M NaCl and 3.5% PEG8000 at pH 6.1 for rAAV9 for a starting sample with 48% full capsids, and (c) 0.6M NaCl, 2.5% PEG8000 at pH 5.7 for rAAV8 for a starting sample with 55% full capsids. The initial capsid concentration was $10^{14}$ vg/ml. TE represents the transduction efficiency measured on day 5 post transduction using Countess II cell counter (Invitrogen).

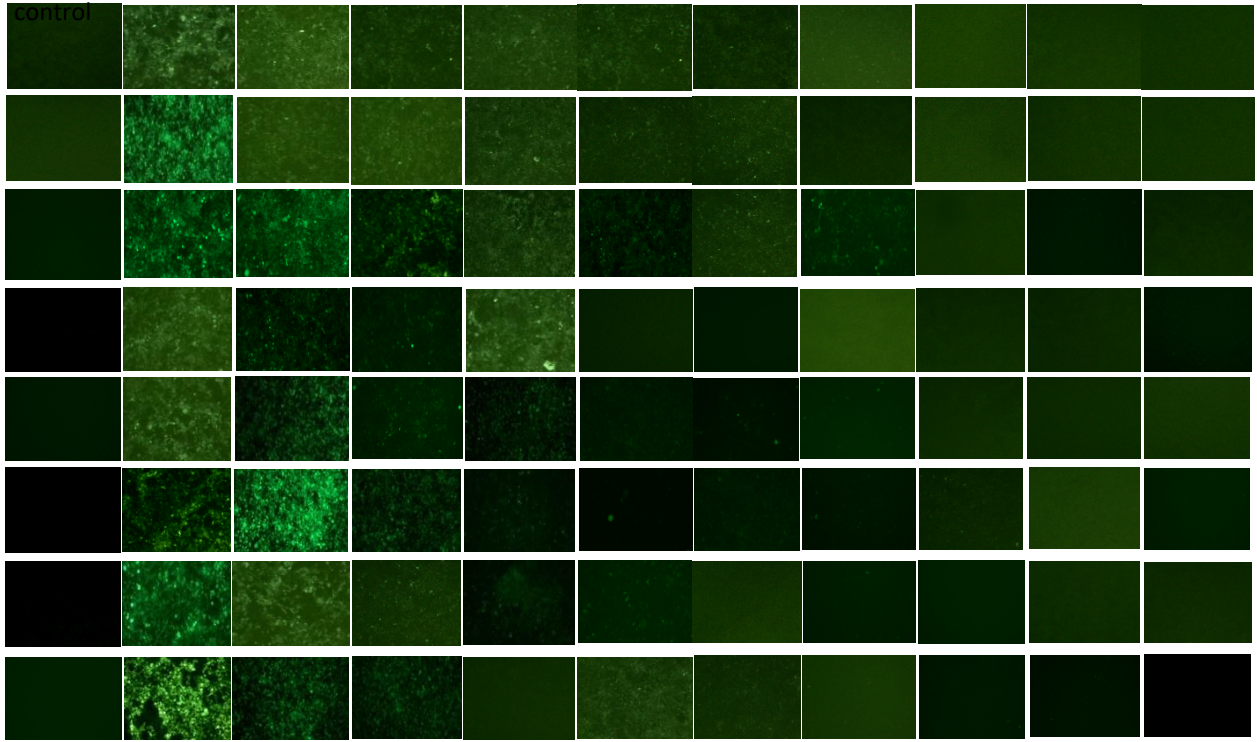

**Fig.17:** An exemplary TCID$_{50}$ experiment showing the gfp fluorescence image 120 h post transduction for transduction of HEK293T adherent cells with full capsids of sf9 produced rAAV5 in crystal stock solution as prepared from the crystals obtained after crystallization in preferential crystallization region of sf9 produced full capsids. For all the serotypes, full capsid's transgene length was 2.46 kbp. Experimental conditions for preferential crystallization of full capsids: 1.4M NaCl and 3.5% PEG8000 at pH5.9 for rAAV5 for a starting sample with 45% full capsids

**Table1:** TCID$_{50}$ results before and after crystallization

|  | rAAV5 | rAAV9 |
|---|---|---|
| Reference sample (before crystallization) | $1.5 \times 10^4 TCID_{50}/0.1mL$ in five days | $1.3 \times 10^4 TCID_{50}/0.1mL$ in five days |
| Crystal sample | $2.5 \times 10^4 TCID_{50}/0.1mL$ in five days | $7.8 \times 10^3 TCID_{50}/0.1mL$ in five days |

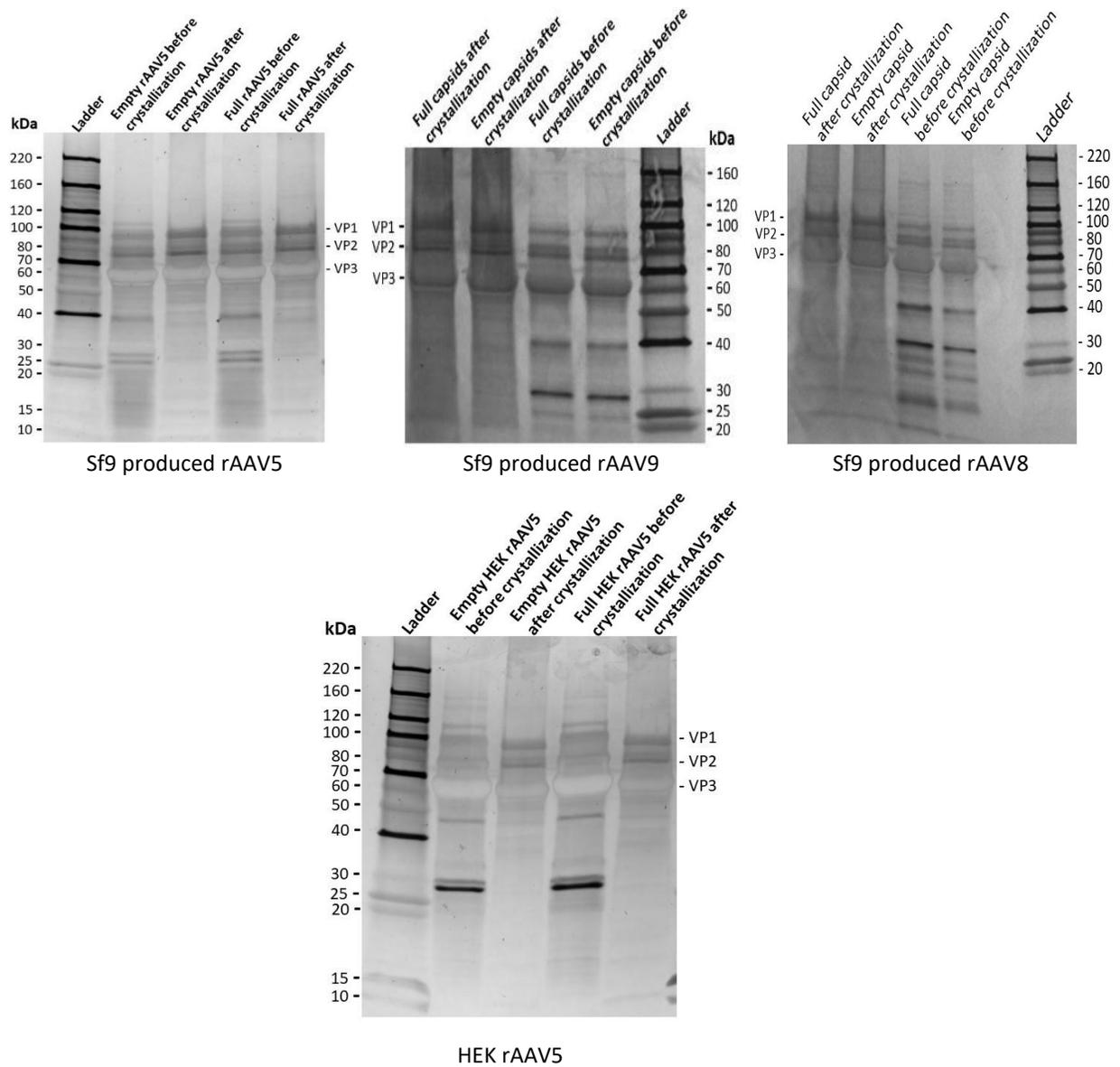

**Fig. 18:** SDS-PAGE gel electrophoresis followed by silver stain results for rAAV capsids before and after preferential crystallization showing the viral capsid proteins, VP1, VP2,and VP3 along with the presence of protein impurities. Experimental conditions for preferential crystallization of full capsids: 3.5% PEG8000, 1.5 M NaCl at pH 5.7 for sf9 produced rAAV5 for a starting sample with 35% full capsids; 4% PEG8000, 1 M NaCl at pH 6.3 for sf9 produced rAAV9 for a starting sample with 24% full capsids; 2.3% PEG8000, 0.3 M NaCl at pH 6.1 for sf9 produced rAAV8 for a starting sample with 37% full capsids, 3% PEG8000, 0.8 M NaCl at pH 5.7 for a starting sample with 30% full capsids of HEK293 cell produced rAAV5. Experimental conditions for preferential crystallization of empty capsids: 2% PEG8000, 0.3 M NaCl at pH 5.7 for sf9 produced rAAV5 for a starting sample with 27% empty capsids; 2.5% PEG8000, 0.6 M NaCl at pH 6.3 for sf9 produced rAAV9 for a starting sample with 40% empty capsids; 4.5% PEG8000, 1.7 M NaCl at pH 6.1 for sf9 produced rAAV8 for a starting sample with 37% empty capsids, 1% PEG8000, 0.8 M NaCl at pH 6.3 for a starting sample with 23% empty capsids of HEK293 cell produced rAAV5.

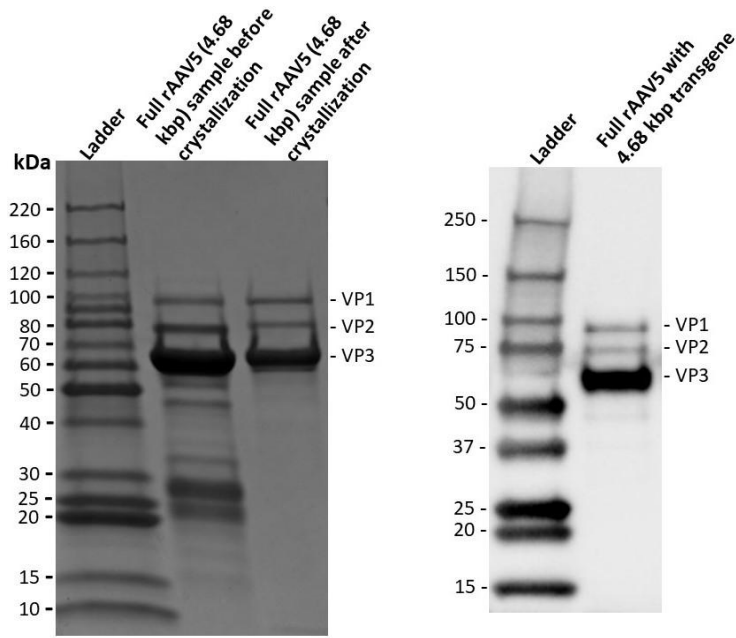

**Fig.19:** SDS-PAGE gel electrophoresis results (left) for full capsids of sf9 produced rAAV5 with transgene length 4.68 kbp, showing the viral capsid proteins, VP1, VP2, and VP3, along with the protein components before and after the preferential crystallization. Western blot results (right) of sample before crystallization identifying only viral protein components and confirming that the low molecular weight protein bands are protein impurities from cell lysis. Experimental condition: 2.83% PEG8000, 1.25 M NaCl at pH 6.3 for a starting sample with 19% full capsids.

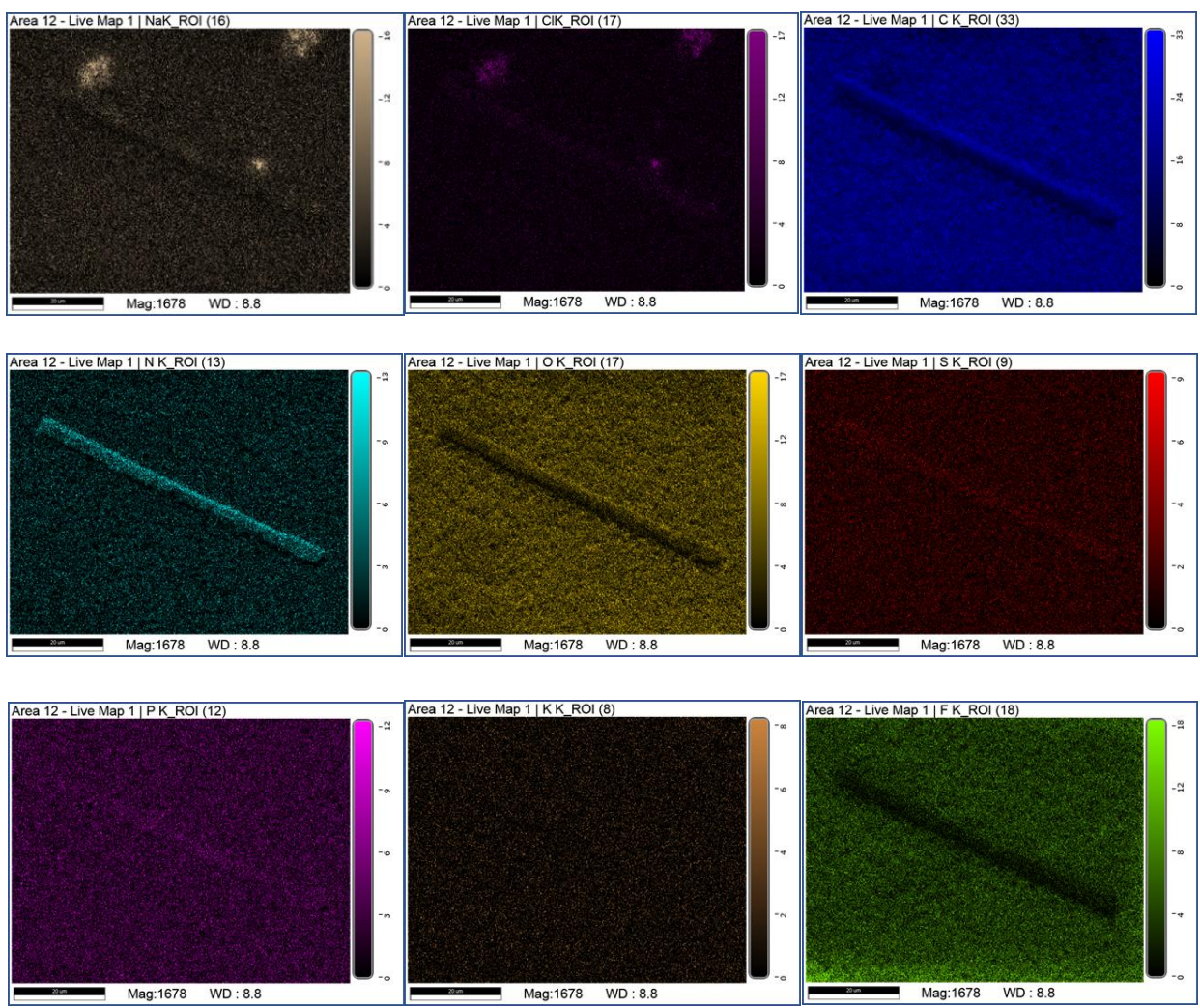

**Fig.20:** Elemental mapping of a crystal of sf9 produced rAAV5 on membrane surface using EDAX (energy dispersive X-ray). Row1: mapping of Na (left), Cl (middle), and C (right); Row2: mapping of N (left), O (middle), and S (right); Row3: mapping of P (left), K (middle), and F (right). Crystal was obtained after crystallization in preferential crystallization region of full capsids. Experimental condition: 3% PEG8000, 1.4 M NaCl, pH 5.7 for rAAV5 for a starting sample with 26% full capsids

**Table 2:** Elemental composition as determined by EDAX analysis in three different locations on crystals of sf9 produced rAAV5, rAAV9, and rAAV8 as obtained by crystallization in preferential crystallization region of full capsids. Preferential crystallization conditions for full capsids: 3% PEG8000, 1.4 M NaCl, pH 5.7 for rAAV5 for a starting sample with 26% full capsids; 4% PEG8000, 1 M NaCl, pH 5.7 for rAA9 for a starting sample with 23% full capsids; 2.5% PEG8000, 0.3 M NaCl and pH 6.1 for rAAV8 for a starting sample with 35% full capsids.

| Area Elements | rAAV5 (atomic %) | | | rAAV9 (atomic %) | | | rAAV8 (atomic%) | | |
|---|---|---|---|---|---|---|---|---|---|
| | 1 | 2 | 3 | 1 | 2 | 3 | 1 | 2 | 3 |
| C K | 56.50 | 56.13 | 55.82 | 56.50 | 54.04 | 55.37 | 54.53 | 55.46 | 54.59 |
| N K | 12.14 | 13.21 | 13.14 | 11.74 | 13.96 | 13.05 | 13.80 | 14.42 | 13.74 |
| O K | 23.22 | 23.27 | 22.92 | 23.44 | 24.23 | 24.10 | 24.10 | 22.84 | 23.56 |
| NaK | 1.01 | 0.83 | 0.95 | 1.13 | 1.02 | 0.86 | 0.93 | 0.63 | 0.89 |
| P K | 3.28 | 3.21 | 3.01 | 3.17 | 2.76 | 3.14 | 2.88 | 3.17 | 2.96 |
| S K | 2.94 | 2.53 | 3.28 | 3.00 | 3.15 | 2.64 | 2.89 | 2.97 | 3.14 |
| Cl K | 0.91 | 0.82 | 0.88 | 1.02 | 0.84 | 0.84 | 0.87 | 0.51 | 1.12 |

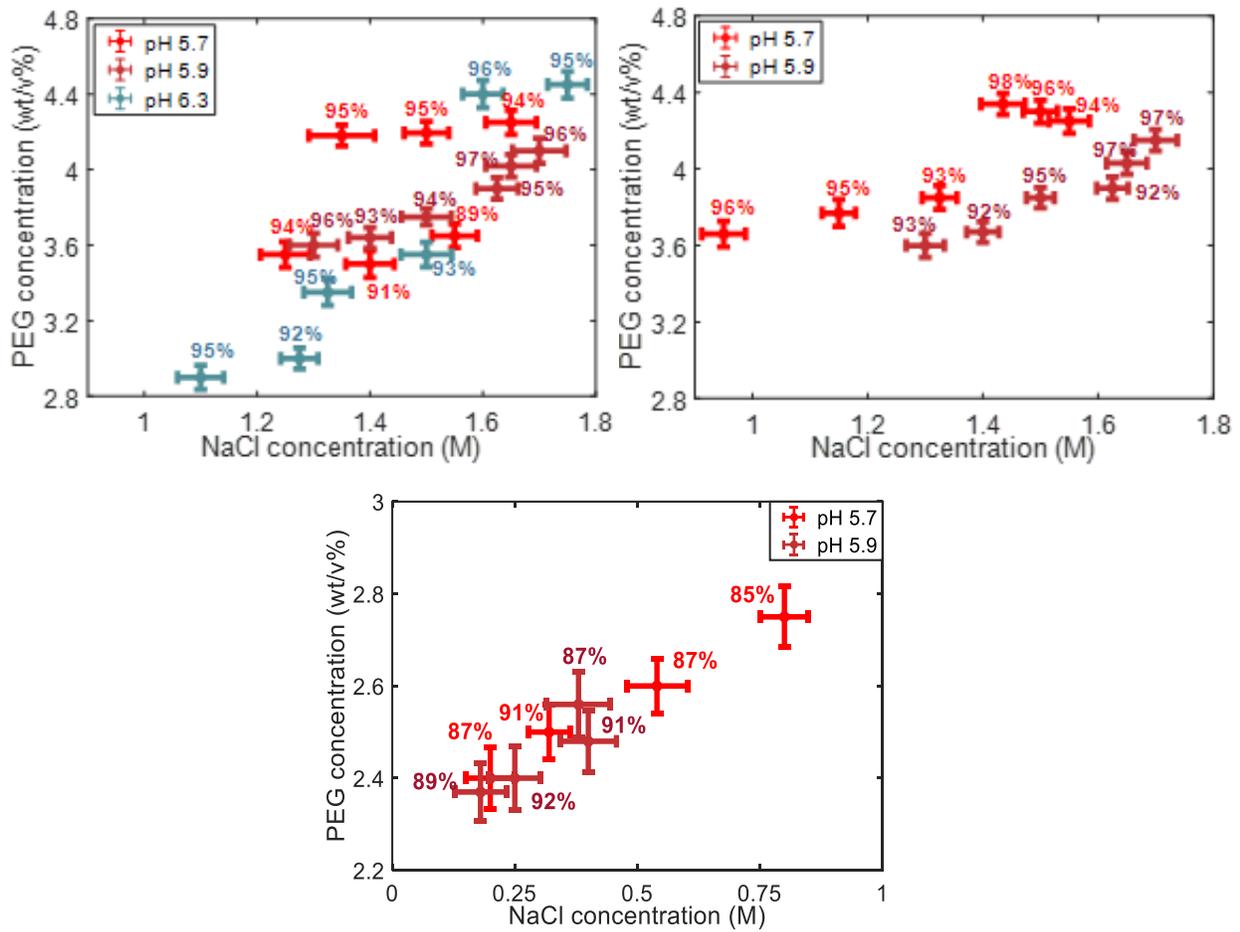

**Fig.21:** Yield of preferential crystallization experiments performed at different crystallization conditions in the preferential crystallization region for full capsids of sf9 produced rAAV5 (left), rAAV9 (middle), and rAAV8 (right). Starting rAAV concentration is $10^{14}$ vg/mL for all serotypes. Starting sample composition: 29% full capsids (rAAV5), 42% full capsids (rAAV9), and 45% full capsids (rAAV8).